# Atmospheric parameters and chemical abundances within 100 pc. A sample of G, K, and M main-sequence stars.

Ricardo López-Valdivia,[1]* Lucía Adame,[1] Eduardo Zagala Lagunas,[2] Carlos G. Román-Zúñiga,[1] Jesús Hernández,[1] Edilberto Sánchez,[1] José G. Fernández-Trincado,[3] Leticia Carigi,[4] Marina Kounkel,[5] Richard R. Lane,[6] Keivan G. Stassun,[7] Sandro Villanova[8]

[1] *Instituto de Astronomía, Universidad Nacional Autónoma de México, Ap. 106, 22800 Ensenada, BC, México*
[2] *Facultad de Ciencias, Universidad Autónoma de Baja California, 22860 Ensenada, B. C., México*
[3] *Instituto de Astronomía, Universidad Católica del Norte, Av. Angamos 0610, Antofagasta, Chile*
[4] *Instituto de Astronomía, Universidad Nacional Autónoma de México, Ap. 70-264, 04510 CDMX, México*
[5] *Department of Physics and Astronomy, University of North Florida, 1 UNF Dr., Jacksonville, FL 32224, USA*
[6] *Universidad Bernardo O'Higgins, Centro de Investigación en Astronomía, Avenida Viel 1497, Santiago, Chile*
[7] *Vanderbilt University, Department of Physics and Astronomy, VU Station 1807, Nashville, TN 37235, USA*
[8] *Universidad de Concepción, Departamento de Astronomía, Casilla 160-C, Concepción, Chile*

**ABSTRACT**
To date, we have access to enormous inventories of stellar spectra that allow the extraction of atmospheric parameters and chemical abundances essential in stellar studies. However, characterizing such a large amount of data is complex and requires a good understanding of the studied object to ensure reliable and homogeneous results. In this study, we present a methodology to measure homogenously the basic atmospheric parameters and detailed chemical abundances of over 1600 thin disk main-sequence stars in the 100 pc solar neighborhood, using APOGEE-2 infrared spectra. We employed the code `tonalli` to determine the atmospheric parameters using a prior on log $g$. The log $g$ prior in `tonalli` implies an understanding of the treated population and helps to find physically coherent answers. Our atmospheric parameters agree within the typical uncertainties (100 K in $T_{\rm eff}$, 0.15 dex in log $g$ and [M/H]) with previous estimations of ASPCAP and Gaia DR3. We use our temperatures to determine a new infrared color-temperature sequence, in good agreement with previous works, that can be used for any main-sequence star. Additionally, we used the `BACCHUS` code to determine the abundances of Mg, Al, Si, Ca, and Fe in our sample. The five elements (Mg, Al, Si, Ca, Fe) studied have an abundance distribution centered around slightly sub-solar values, in agreement with previous results for the solar neighborhood. The over 1600 main-sequence stars' atmospheric parameters and chemical abundances presented here are useful in follow-up studies of the solar neighborhood or as a training set for data-driven methods.

## 1 INTRODUCTION

In the last decades, thanks to large-area spectroscopic surveys like the The RAdial Velocity Experiment (RAVE, Steinmetz 2003; Guiglion et al. 2020), the Large Sky Area Multi-Object Fiber Spectroscopic Telescope (LAMOST, Cui et al. 2012; Zhao et al. 2012), the Gaia-ESO (Gilmore et al. 2012), the GALactic Archeology with Hermes (GALAH, De Silva et al. 2015; Martell et al. 2017), and the Sloan Digital Sky Survey (SDSS) APOGEE-2 (Majewski et al. 2017; Wilson et al. 2019), we have access to enormous inventories of stellar spectra of stellar populations with different spectral types and at different evolutive stages. These and other surveys are designed to provide uniform-quality data with sufficient signal-to-noise (SNR) and spectral resolution to allow precise spectral classifications and the extraction of reliable spectral parameters essential to characterize the stellar ecosystems of the Milky Way.

Characterizing this amount of data is a complicated task and the surveys often solve it by developing general pipelines that, on average, work well over all types of stars. The estimation of the basic atmospheric parameters, such as effective temperature ($T_{\rm eff}$), surface gravity (log $g$), and overall metallicity ([M/H]), often relies on the comparison between the observed and synthetic spectra. Although, over the years the synthetic spectra grids have improved giving us a larger and more regular coverage of the parameter space (e.g., Gustafsson et al. 2008; Husser et al. 2013; Allard et al. 2012), the reliability of the stellar parameters obtained varies according to the spectral type or evolutionary stage of the analyzed object.

These discrepancies are often solved by introducing *a posteriori* calibration to match the solar stellar parameters of another well-known star (e.g., Arcturus). For instance, the SDSS pipeline in its different released versions applied a posterior calibration to their spectroscopic temperatures, gravities, and also chemical abundances. In SDSS data release 16 (DR16; Jönsson et al. 2020) the spectroscopic temperatures are calibrated to the photometric scale of

* E-mail: rlopezv@astro.unam.mx

© 2024 The Authors

González Hernández & Bonifacio (2009), while for gravities, DR16 implemented different calibration categories according to the evolutionary stage of the star based on asteroseismology and isochrones. The methodology to calibrate log *g* changed in data release 17 (DR17; Abdurro'uf et al. 2022) to a neural network approach that eliminates small discontinuities introduced by the categorization. *A posteriori* calibration risks to provide physically incorrect parameters, which can complicate or bias subsequent analyses.

Recently, to characterize large stellar samples some studies have used data-driven techniques (e.g., Ness et al. 2015; Sprague et al. 2022; Sizemore et al. 2024). Under this approach, a set of well-known empirical spectra is used as a training set of a prediction system to infer the stellar properties. The training set is of great importance as it must be a reliable sample covering the complete parameter space of interest. Of course, the data-driven techniques are not able to infer properties outside of the limits of their training sets.

Another option is the use of *a priori* approach to guide the stellar parameter determination analysis. As *a priori* constraint relies on previous knowledge, they should have more physical sense than some *a posteriori* calibrations. In this paper, we determined the atmospheric parameters and the chemical abundances for 1600+ main-sequence stars using an *a priori* approach described in the next sections.

Individual chemical abundances provide information not only on stellar astrophysics but also on the formation and chemical evolution of the Galaxy (e.g., Tinsley & Larson 1978; Bensby et al. 2014; Goswami & Prantzos 2000; Neves et al. 2009; Carigi et al. 2019; Palicio et al. 2023). Present-day chemical abundances of the stars are also helpful for Galactic archaeology, which is an approach applicable to the Milky Way and other galaxies (e.g., Kobayashi 2016) where the analysis of abundance patterns informs us on the star formation history of the host galaxy. On the other hand, the atmospheric parameters and chemical abundances are also of interest in the exoplanets field (e.g., Ramírez et al. 2009; Yana Galarza et al. 2021; Cowley & Yüce 2022; Pignatari et al. 2023) either to identify host stars through their chemistry or to derive other stellar parameters such as mass and radius that in turn serve to characterize the exoplanet properties. Furthermore, by combining the atmospheric parameters and chemical abundances with orbital information, we can do chemo-chronological analyses (e.g., da Silva et al. 2012, 2023). In this work we present, test, and characterize a methodology that produces atmospheric parameters and chemical abundances consistent with previous determinations without the need for a posterior calibration. We seek to implement these methods in different stellar populations, especially low-mass pre-main sequence stars.

We organized this paper as follows: in Section 2, we present our stellar sample and detail the infrared spectroscopic data obtained from the SDSS phase IV. In Section 3 and 4, we describe the atmospheric parameter and chemical abundance determination process, respectively, while in Section 5, we summarize our main results.

## 2 SPECTROSCOPIC DATA

The Apache Point Observatory Galactic Evolution Experiment (APOGEE) and APOGEE-2 programs in the third and fourth phases of SDSS used two state-of-the-art high-resolution (R~22 500) fiber spectrographs with the capacity for up to 300 simultaneous fiber observations. The dispersed light from the fibers is collected by bench spectrographs and divided into three near-IR detectors: blue 15145–15810 Å, green 15860–16430 Å, and red 16480–16950 Å. One of the APOGEE spectrographs is installed in the Northern Hemisphere (APOGEE-N), on the 2.5m Sloan Foundation telescope at Apache Point Observatory (Gunn et al. 2006), and the other one is located in the Southern Hemisphere (APOGEE-S), on the 2.5m Du Pont telescope at Las Campanas Observatory (Bowen & Vaughan 1973). Both instruments are essentially identical (Wilson et al. 2019), ensuring homogeneity in the data. For this paper, we used publicly available epoch-combined 1D spectra from the APOGEE Data Release 17 (DR17, Abdurro'uf et al. 2022).

### 2.1 Sample selection

We first cross-matched the APOGEE DR17, the Gaia DR3 (Gaia Collaboration et al. 2023a), and the Bailer-Jones et al. (2021) catalogs. A total of 605244 sources are common to the three catalogs. Bailer-Jones et al. (2021) through a probabilistic approach and Gaia's parallaxes infer two types of distances namely: geometric and photogeometric. The geometric approach uses the parallax with a direction-dependent prior on distance. While the photogeometric, additionally uses the color and apparent magnitude of the star. This second approach takes the advantage of that stars of a given color have a restricted range of probable absolute magnitudes (plus extinction). In this study, we used the geometric distances to select those stars closer than 100 pc and Gaia Renormalised Unit Weight Error (RUWE) less than 1.4. The RUWE parameter measures the goodness of the astrometric fit, and it is expected to be about 1.0 for a single star. Values of RUWE greater than 1.4 might indicate that the source is non-single or that the astrometric solution has an issue[1]. Constraining our sample to RUWE values lower than 1.4 eliminates most of the binary contamination (Gaia Collaboration et al. 2023b) and guarantees that we use the best astrometric solutions. The distance cut was imposed at 100 pc as we expect zero (or close to zero) extinction in the main-sequence stars. Also, we expected to obtain a mean SNR in the APOGEE-2 spectra high enough to extract reliable atmospheric parameters.[2] We found that 2647 stars met the distance and RUWE criteria. We then cross-matched this list with the 2MASS catalog (Cutri et al. 2003) to obtain their *J* and *H* band photometry, which together with Gaia's *G*, *BP*, and *RP* photometry, allowed us to construct a (*BP* − *RP*) vs (*G* − *J*) color- color diagram of the sample. The selected stars form a tight sequence in this color space, to which we fitted a third-order polynomial. Applying a $\sigma$ clipping criterium, we rejected stars that differed more than 0.1 mags of the fitted line. Using the (*BP-RP*) vs *G* absolute magnitude diagram, we also identified and removed a few stars not located in the main sequence. Here, we also applied a $\sigma$ clipping using the median and the median absolute deviation of colors in several bins of absolute magnitude. Finally, we used the proper motions (in RA and Dec) of Gaia Collaboration et al. (2021), the geometric distance of Bailer-Jones et al. (2021), and the radial velocity of the APOGEE DR17 to calculated the Galactic space-velocity components, relative to the local standard of rest (LSR; Coşkunoğlu et al. 2011): $U_{LSR}$, $V_{LSR}$, and $W_{LSR}$ and the total velocity $v_{tot} = (U_{LSR}^2 + V_{LSR}^2 + W_{LSR}^2)^{1/2}$. We then selected those stars with $v_{tot} \leq 50$ km s$^{-1}$ as they are most probably to belong to the thin disc (e.g., Bensby et al. 2014). After all the mentioned cuts, our final sample comprises 1662 main-sequence stars.

---

[1] see Gaia documentation

[2] About 98% of our final sample has a mean SNR ≥ 50.

**Table 1.** Parameter space used in this work. These intervals are set by the MARCS synthetic grid used (Jönsson et al. 2020).

| Parameter | Range | Step |
|---|---|---|
| $T_{eff}$ | 3000 — 8000 K | 100 K for $T_{eff} \leq 4000$ K; 250 K otherwise |
| $\log g$ | 0.0 — 5.5 dex | 0.5 dex |
| [M/H] | −2.50 — 1.00 dex | 0.25 dex |
| [α/Fe] | −0.75 — 1.00 dex | 0.25 dex |

## 3 ATMOSPHERIC PARAMETERS

We determined the effective temperature ($T_{eff}$), surface gravity ($\log g$), overall metallicity ([M/H]), α elements enhancement ([α/Fe])[3], and projected rotational velocity ($v \sin i$) with the `tonalli` code (Adame et al. in prep). In short, `tonalli` is a multi-dimensional heuristic interpolation code based on the Asexual Genetic Algorithm (AGA) by Cantó, Curiel & Martínez-Gómez (2009), developed to determine accurate atmospheric parameters mainly from spectroscopic data for a diversity of stellar populations, such as, pre-main-sequence and main-sequence low-mass stars. `tonalli` randomly selects a sample of stellar parameters from a grid of stellar atmosphere models using user-selected intervals or grid limits. Then, synthetic spectra are interpolated at the selected parameters, and each of these random-selected synthetic spectra is compared to the epoch-combined 1D observed spectrum through a $\chi^2$ statistic. Those stellar parameters that produce the synthetic spectra with the lower $\chi^2$ values are selected as progenitors to create the following generation of synthetic spectra, which are created by randomly selecting spectra from a decreasing vicinity around each progenitor. The process is repeated with the search vicinity around the progenitors decreasing in size each generation until the stop criteria are achieved. For our analysis, we employed the theoretical grid of Jönsson et al. (2020), which was specially developed to work with APOGEE data and is based on MARCS models (Gustafsson et al. 2008). The `tonalli` code can work on other synthetic grids, and the parameter space can be chosen by the user depending on the problem to investigate. In Table 1, we report the space of parameters that we use in this study We only allowed `tonalli` to interpolate models with $v \sin i \leq 50$ km s$^{-1}$, as our sample is composed of main-sequence stars, and we are not expecting a significant fraction of stars with velocities higher than the resolution limit of our data, which is ∼13 km s$^{-1}$ (Cottaar et al. 2014; Serna et al. 2021).

### 3.1 $\log g$ prior

The $\log g$ is the most complicated parameter to determine from stellar spectra, as its effects on the line profiles might be similar to those caused by other effects like stellar rotation or abundance. We found evidence of model degeneracies in `tonalli` that resulted in an unsuitable combination of stellar parameters due to inaccurate $\log g$ estimates. To get around this problem, our $\log g$ space depends on a Gaussian prior defined for each star, as explained in the following paragraphs.

The most recent version of `tonalli` includes a photometric module called `tepitzin` (Adame et al. in prep) that provides informed priors for $T_{eff}$, $\log g$, or [M/H] if the user needed. `tepitzin` requires the 2MASS *J*, *H*, and *K* magnitudes and the Gaia *G*, *BP*, and *RP*, as well as the parallax (or distance) to the star. With this information, `tepitzin` also implements the AGA, to look for the best combination of $T_{eff}$, $\log g$, [M/H], and stellar age that minimizes the differences between the observed absolute magnitudes and the absolute magnitudes of a set of evolutionary models, in this case, the isochrones of PARSEC (Marigo et al. 2017). We used the PARSEC isochrones with ages between $10^8$ and $10^{10}$ years, while the metallicities span from -2.0 to 0.5 dex. This way, `tepitzin` can provide us with a preliminary estimation of photometric $T_{eff}$, $\log g$, [M/H], and age, assuming the visual extinction towards the star is $A_V = 0$. We note that `tepitzin` is a module that can be used as a standalone code to infer photometric parameters, but it was constructed to be a complement to `tonalli` to improve the derivation of the spectroscopic parameters, more details on `tepitzin` are available in Adame et al. (in prep).

In this study, we used the $\log g$ photometric value of `tepitzin` to construct a prior for $\log g$ in `tonalli`. This prior is defined as a Gaussian kernel centered on the photometric $\log g$ with a standard deviation of 0.5 dex[4]. The $\log g$ prior depends on the photometric colors of the star, and it will be different for each star in our sample.

### 3.2 `tonalli` results

We used `tepitzin` to produce priors for each run of `tonalli` in our sample of spectra. In its last step, `tonalli` computes confidence intervals for each parameter, by repeating the fitting process for the last generation *n* times. For each star, the results generated from the complete process are saved into a file that contains all necessary statistics.

To determine the atmospheric parameters of our sample, we performed *n* = 30 repetitions in `tonalli` as this number of repetitions is enough to guarantee the convergence of the code (Adame et. al. in prep). We use the median (50 percentile) of the posterior distributions as the estimator for the parameter analyzed, while for the uncertainties, we took the interquartile range[5]. We report in Table 5 the atmospheric parameters found for our sample. We discard from our analysis those stars that have an interquartile range in $T_{eff}$ larger than 500K and in $\log g$ larger than 0.3 dex, as those determinations are likely inaccurate. This limited sample comprises 1653 stars that we will analyze in the following sections.

In Figure 1, we show the Kiel diagram ($\log g$ vs. $T_{eff}$) for our sample. We can see that most of our stars in this diagram are located in the expected regions, as defined by the modeling isochrones. The good agreement between our resultant parameters and the model isochrones, indicates a correct behavior of the $\log g$ prior, and justifies not using a posterior calibration.. Our sample spans from 3000 to ∼6000 K, which corresponds to spectral types between ∼F9 and M5 (Pecaut & Mamajek 2013).

#### 3.2.1 Comparing `tonalli` parameters with literature

The natural comparison to our stellar parameters is the APOGEE Stellar Parameters and Chemical Abundances Pipeline (ASPCAP; García Pérez et al. 2016) DR16 version (Jönsson et al. 2020) values, as we are using the same synthetic spectra and similar (if not the same) observed spectra in `tonalli`. The ASPCAP Pipeline uses a minimum $\chi^2$ statistic to compare observed APOGEE spectra to a set of model libraries in a two-step process, to determine stellar parameters and abundances. In the first step a set of $T_{\rm eff}$, $\log g$, [M/H], $v \sin i$, and microturbulence velocity is fitted to the entire APOGEE spectrum. Additionally, the abundance of C, N, and $\alpha$-elements can also vary during this step. Once the stellar parameters are found, ASPCAP determines individual abundances through the fitting of spectral windows carefully selected for this purpose.

---

[3] The [α/Fe] includes changes in O, Ne, Mg, Si, S, Ar, Ca, and Ti.

[4] This value is set by the user. In our case, a value of 0.5 is enough to have a correct mapping of the $\log g$ space.

[5] Defined as the difference between the 75 and 25 percentile of the posterior distribution.

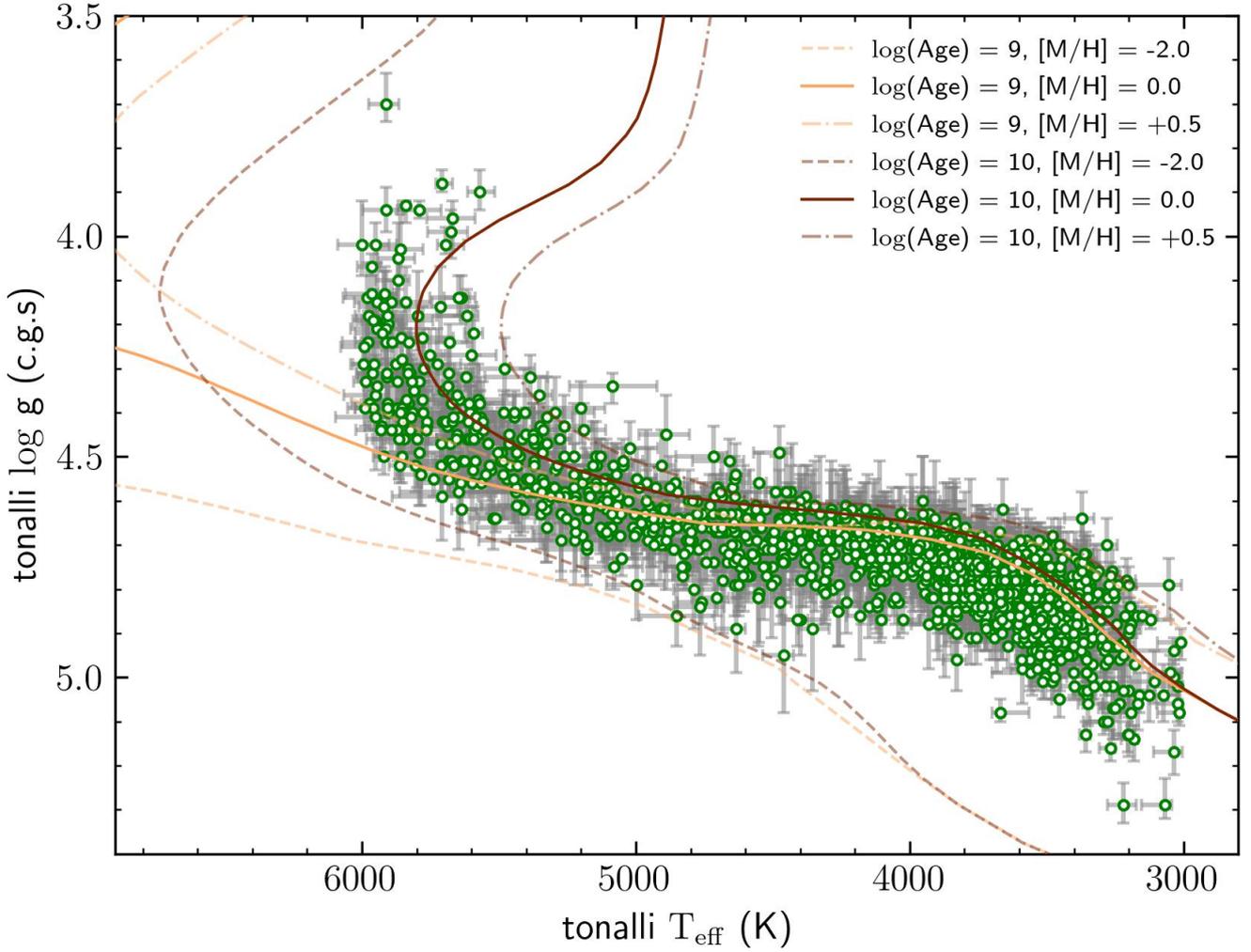

**Figure 1.** Kiel diagram for the 1653 stars of our sample obtained with `tonalli`. The error bars represent the percentiles 25 and 75. As a reference we include, as color lines, evolutionary models of Marigo et al. (2017) for log(Age) of 9 and 10 at three different metallicity values.

As we mentioned before, DR16 made a posterior calibration to its $T_{\rm eff}$ and $\log g$ values, resulting in two different sets of parameters: the *spec*, those without any calibration, and the *calibrated*, which are the *spec* values after the implemented calibration. In the following, we will focus on the calibrated DR16 atmospheric parameters, but we extended our comparison in Appendix A to the *spec* DR16 parameters and the *spec* and calibrated DR17 estimations. In Figure 2, we show the Kiel diagram that corresponds to DR16 $T_{\rm eff}$ and $\log g$ before and after calibration.

The corresponding Kiel diagram for the *spec* DR16 values (left panel in Fig. 2) did not populate, according to the stellar ages expected for our sample, a reasonable region of this diagram for stars with $T_{\rm eff}$ lower than ~4800 K. The posterior calibration does improve the atmospheric parameters but only partially, as now stars with $T_{\rm eff}$ lower than ~4000 K, appear to have a strong correlation between $T_{\rm eff}$, $\log g$, and [M/H] as the coolest stars have higher $\log g$ and lower [M/H] values, according to the isochrones of PARSEC.

Issues like those present in the *spec* DR16 atmospheric parameters are somehow expected, as the pipeline used to get the parameters is optimized to work with giant stars. Although the ASPCAP pipeline has improved its atmospheric parameter determination for different stellar populations (e.g., giants, main-sequence, red giant branch, etc.) over the years, the DR16 calibrated atmospheric parameters obtained for main-sequence stars still need to be taken with caution, especially for coolest temperatures.

Additionally to the DR16 values, we compare our `tonalli` determinations to Apogee Net III (ANet III; Sizemore et al. 2024), and Gaia DR3 (Gaia Collaboration et al. 2023a).

ANet III (Sizemore et al. 2024), and its predecessor (Olney et al. 2020; Sprague et al. 2022), use a deep convolutional neural network to predict $T_{\rm eff}$, $\log g$, and [M/H] based on previously derived parameters (or "labels") collected for APOGEE data and other stars in diverse studies. In its first version, Olney et al. (2020) used labels defined for red giants and stars with $T_{\rm eff} > 4200$ K from "The Payne" pipeline (Ting et al. 2019), and parameters derived from photometric relations for M and K main-sequence dwarfs as well as pre-main-sequence stars to train the neural network. For the second version of Apogee Net, Sprague et al. (2022) complemented the training set with a collection of literature $T_{eff}$ and log $g$ values that allow a consistent parameter determination across the HR diagram (from late M to OB stars).

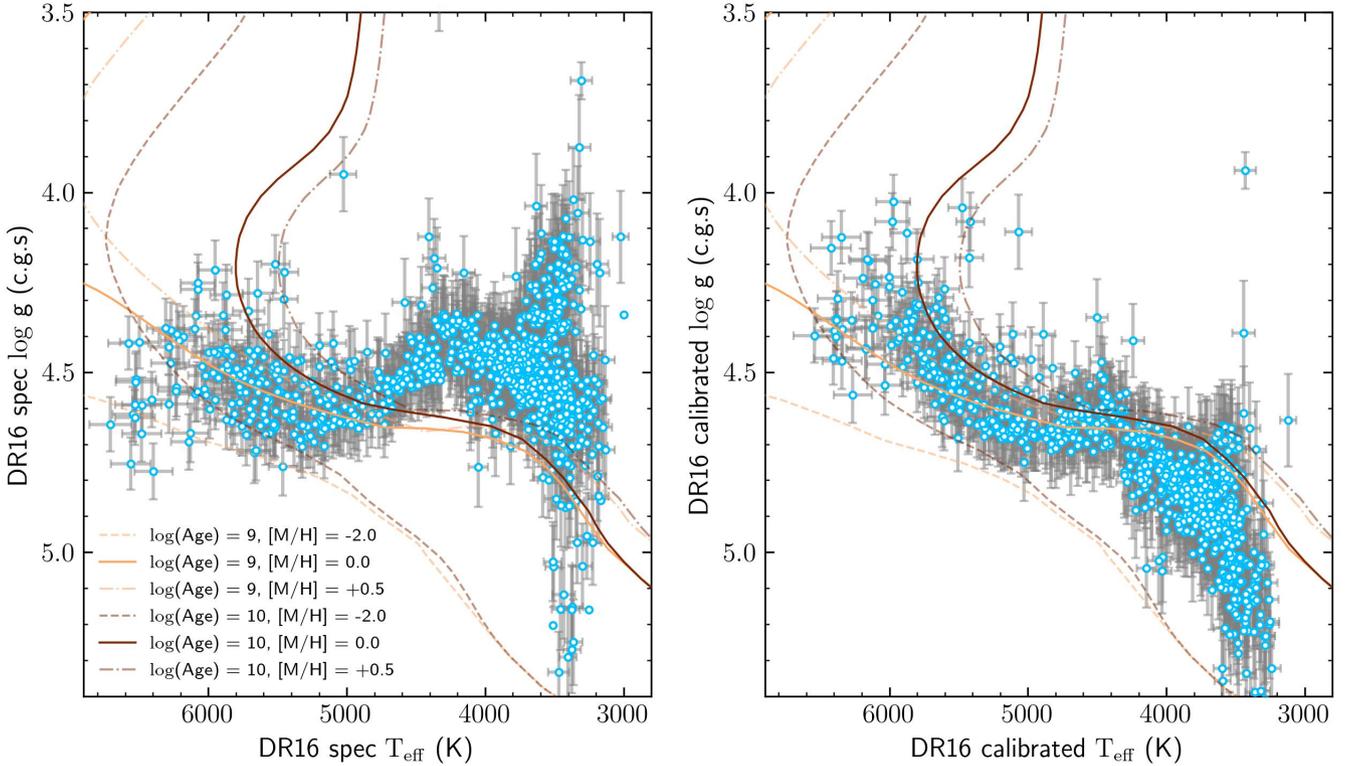

**Figure 2.** Kiel diagram for the 1653 stars of our sample and the atmospheric parameters determined by ASPCAP DR16. On the left panel, we show the APOGEE DR16 spec values (raw parameters determined by the pipeline), while on the right panel, we show the calibrated values. The curves are PARSEC isochrones for the ages indicated in the legend.

Finally, the Gaia DR3 (Gaia Collaboration et al. 2023a) team estimates $T_{eff}$, log $g$, [M/H], radius, absolute magnitude $G$, distance and line-of-sight extinction from low-resolution optical *BP/RP* spectra, apparent *G* magnitude and parallax. They employed a Markov Chain Monte Carlo (MCMC) algorithm in three different steps to derive the parameters mentioned above in a self-consistent manner from four grids of synthetic spectra and a set of PARSEC COLIBRI isochrones (Tang et al. 2014; Chen et al. 2015; Pastorelli et al. 2020).

The studies by Jönsson et al. (2020) and Sprague et al. (2022) employed APOGEE spectra as our work; for this reason, the differences between their estimations and ours might be related more to subtle differences in the methodologies followed and will serve as an internal comparison. On the other hand, the comparison with the Gaia DR3 values will help to assess the `tonalli` parameters and might be useful to characterize systematic offsets.

Figure 3 shows the comparison between our $T_{eff}$ estimates and those by DR16, ANet III, and Gaia DR3. There is good agreement between our $T_{eff}$ estimates and those by the three previous studies up to ~5800 K. Above this interval we found increasingly large discrepancies with the three studies, this is related to the applicability limits of `tonalli` in $T_{eff}$, which according to Adame et al. (in prep), is between 3200 K and ~6000 K when the MARCS theoretical grid is used.

We found that our $T_{eff}$ values are very similar to the calibrated DR16 values, and slightly hotter than those reported by ANet III and Gaia. The median of the differences are ~4 K, ~145 K, and ~55 K concerning ASPCAP DR16, ANet III, and Gaia DR3, respectively. The sequence of residuals to ANet III (center panel of Figure 3) shows a small dip between ~3700 and ~4400 K, where their median temperatures are cooler ~215 K than those determined with `tonalli`. Outside of the dip range, the median $T_{eff}$ differences are ~55 K for hotter stars and ~130 K for cooler stars.

Regarding log $g$, we found that our log $g$ values are more consistent with those reported by ANet III (see Figure 4). The median of the residuals concerning ANet III is ~ −0.002 dex. The agreement with Gaia log $g$ and DR16 is in general also good, with a median of the residuals centered at −0.04 dex and 0.05 dex, respectively. Considering the complication of determining a precise log $g$ value, our derived values for log $g$ can be considered consistent with previously published results.

For [M/H], we found a good agreement with the published values of ASPCAP DR16, shown in the left panel of Figure 5, as the median of the residuals is 0.01±0.15 dex. On the other hand, the residuals for the comparison with ANet III (center panel of Figure 5) show a trend. Most of the [M/H] values of ANet III for our sample gather around a ~-0.07 dex with a small spread ($\sigma = 0.10$ dex), which might be a consequence of the spectroscopic labels used as a training set. Finally, although the comparison with the Gaia DR3 metallicities shows a large dispersion ($\sigma = 0.36$ dex) in the residuals, we found a good agreement (median differences 0.03 dex). The large dispersion found might be related to the fact that the Gaia metallicities were determined from low-resolution spectra.

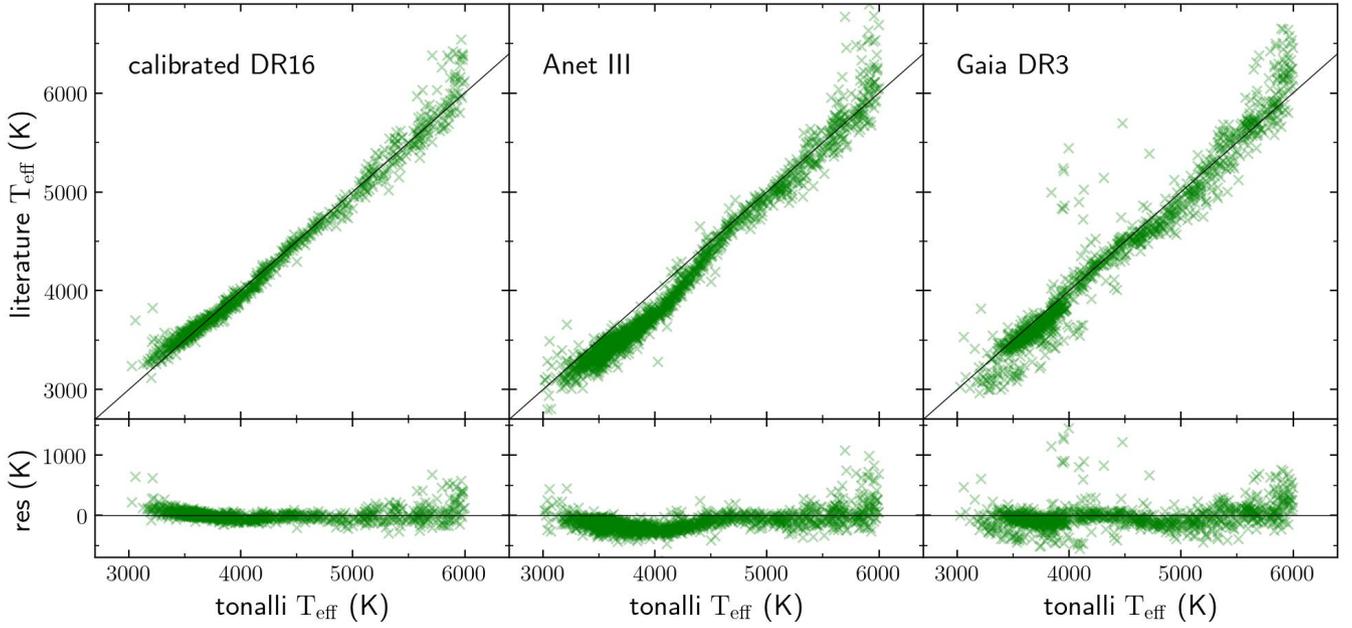

**Figure 3.** Comparison of the `tonalli` $T_{eff}$ with the estimates of ASPCAP DR16 (Jönsson et al. 2020), ANet III (Sizemore et al. 2024) and Gaia DR3 (Gaia Collaboration et al. 2023a). The lower panels are the differences between the literature and our estimates.

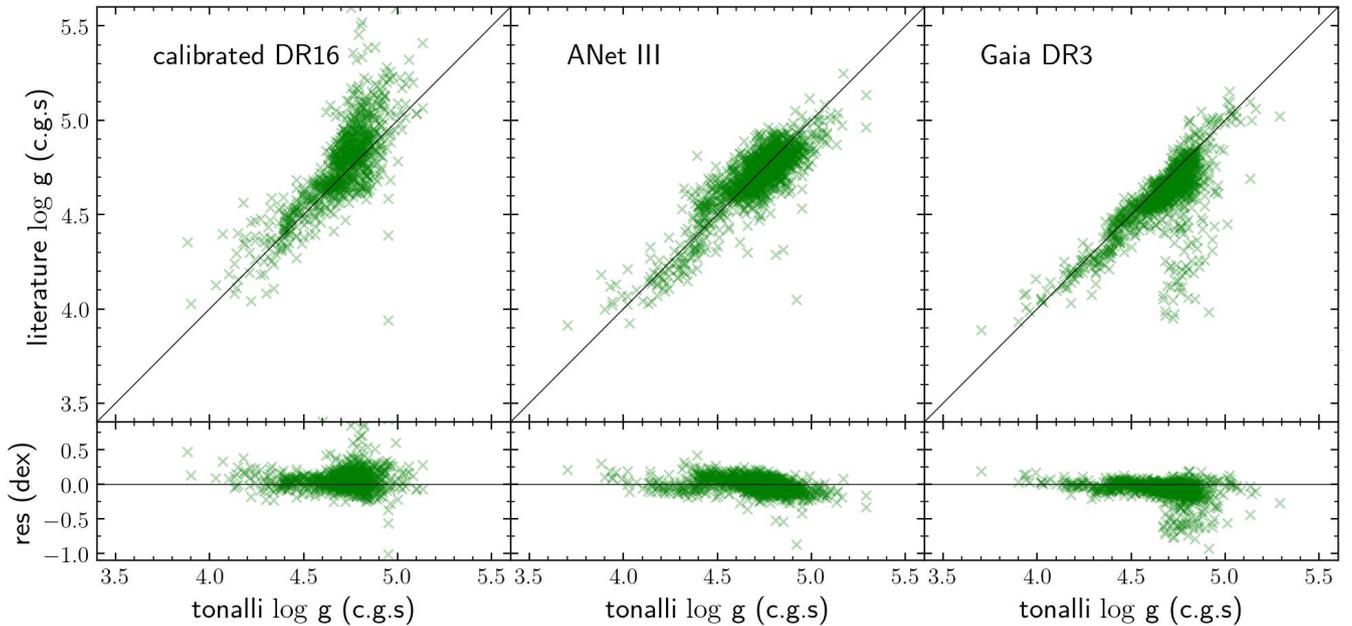

**Figure 4.** Same as Figure 3 but for $\log g$.

To find the best $v \sin i$ value `tonalli` broadens the synthetic spectra with a rotational broadening kernel (Gray 2008). The drawback of this technique is that we cannot determine precise $v \sin i$ for stars with a rotational velocity significantly lower than the spectral resolution limit. Still, this technique recovers acceptable values when rotation dominates the line profile broadening. Our sample has a median $v \sin i$ of ~12 km s$^{-1}$, which compares to the APOGEE-2 resolution limit. This result is expected, as in general main-sequence stars with spectral types G and early K has $v \sin i \geq 5$ km s$^{-1}$, while some late K and M main-sequence stars might rotate faster up to ~15-25 km s$^{-1}$ on average (e.g., Glebocki & Gnacinski 2005; Llorente de Andrés et al. 2021). About 69% (1136 objects) of our sample have a $v \sin i$ value of 13 km s$^{-1}$ or lower, which is a clear sign that `tonalli` is maintaining the rotational broadening basically in its limit, due to the spectral resolution of the data.

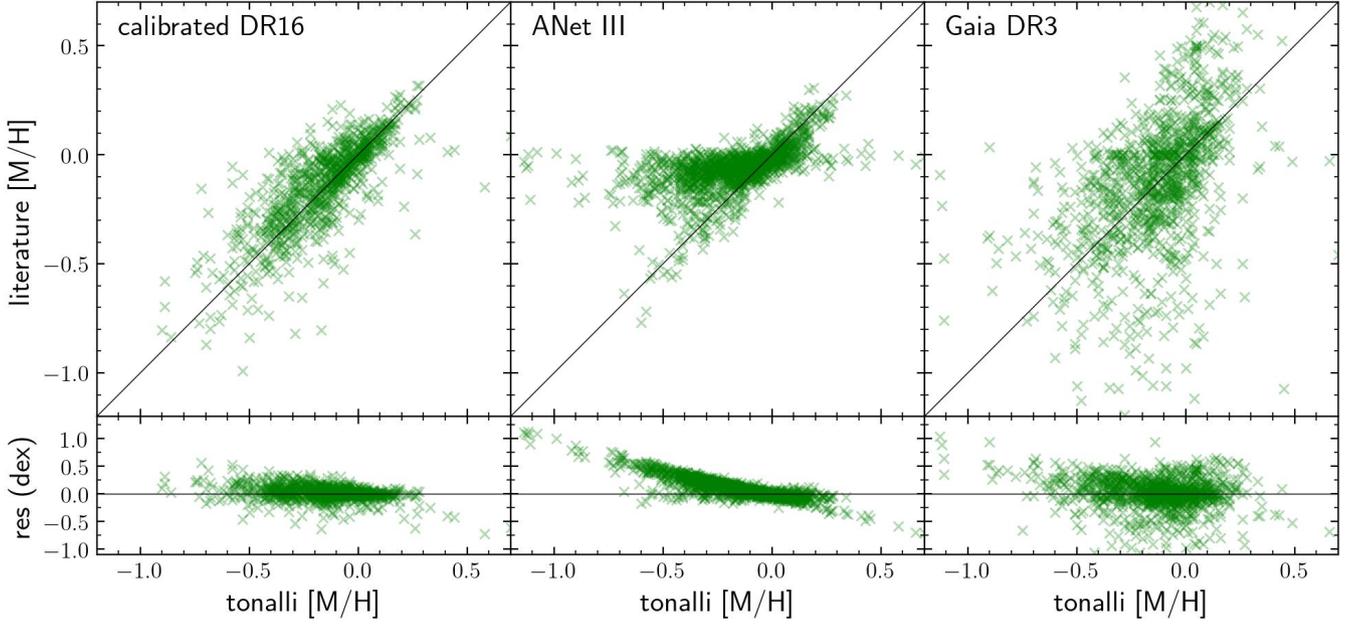

**Figure 5.** Same as Figure 3 but for [M/H].

The characterization we did here with `tonalli` represents a new set of reliable parameters determined from infrared spectra that can be used in diverse astronomical studies. For instance, to derive stellar parameters (e.g., chemical abundances, luminosities, ages), to train neuronal networks or machine learning codes, or to establish relations between stellar parameters. In this sense, we used `tonalli` parameters to establish a color-temperature relation and also to determine the chemical abundances of five elements for our sample in the following sections.

### 3.3 `tonalli` color-temperature relations

We used the 2MASS $H$ and $K$ magnitudes and Gaia $G$, $BP$, and $RP$ to produce optical and infrared color–temperature relations. We binned our data into ten $T_{eff}$ bins and computed the median color to construct six ($J-H$, $H-K$, $G-H$, $BP-H$, $RP-H$, $BP-RP$) color–temperature relations which we report in Table 2 and show in Figure 6.

We compared our resulting color sequences with the well-known intrinsic color relations by Pecaut & Mamajek (2013)[6]. As a reference we also plot, in Figure 6, the theoretical colors of PARSEC (Marigo et al. 2017). Although some differences, we found a good agreement with Pecaut & Mamajek (2013) in all diagrams. The median absolute difference (MD) between the Pecaut & Mamajek (2013) colors and ours span a range from 0.01 mag (for $H-K$) to 0.19 mag (for $BP-H$), which is equivalent to a ~2-5% considering the color ranges of such diagrams. We found comparable results when compared our colors with those of Marigo et al. (2017) for $G-H$, $BP-H$, $RP-H$, and $BP-RP$, as the PARSEC color-temperature relations are similar to that of Pecaut & Mamajek (2013). For the $J-H$ and $H-K$ diagrams the PARSEC relations differ more (especially in $J-H$), but still the agreement is acceptable. We also include a polynomial fit to the Gaia DR3 $T_{eff}$ values and show it in the $BP$–$RP$ diagram (lower right panel of Figure 6). The agreement achieved in this diagram is excellent as we found an MD of 0.06 mag, while for the PARSEC and Pecaut & Mamajek (2013) relations is of ~0.12 mag.

---

[6] We used the relations as published in the website https://www.pas.rochester.edu/~emamajek/EEM_dwarf_UBVIJHK_colors_Teff.txt as of February 2024.

**Table 2.** Optical/Infrared color-temperature relations found in this work.

| log T$_{eff}$ | $J-H$ | $H-K$ | $G-H$ | $BP-H$ | $RP-H$ | $BP-RP$ |
|---|---|---|---|---|---|---|
| 3.49 | 0.588 | 0.275 | 3.750 | 5.590 | 2.457 | 3.117 |
| 3.52 | 0.588 | 0.258 | 3.510 | 5.110 | 2.291 | 2.807 |
| 3.55 | 0.598 | 0.239 | 3.280 | 4.639 | 2.132 | 2.499 |
| 3.58 | 0.626 | 0.218 | 3.042 | 4.168 | 1.988 | 2.180 |
| 3.61 | 0.648 | 0.175 | 2.779 | 3.691 | 1.831 | 1.853 |
| 3.64 | 0.627 | 0.147 | 2.503 | 3.242 | 1.658 | 1.587 |
| 3.67 | 0.548 | 0.116 | 2.182 | 2.777 | 1.443 | 1.335 |
| 3.70 | 0.452 | 0.103 | 1.842 | 2.318 | 1.205 | 1.113 |
| 3.73 | 0.344 | 0.077 | 1.519 | 1.896 | 0.971 | 0.917 |
| 3.76 | 0.248 | 0.063 | 1.230 | 1.524 | 0.759 | 0.748 |

The `tonalli` temperature–color sequences that we present here are as good as previous scales (e.g., Pecaut & Mamajek 2013; Marigo et al. 2017), and they represent a newer alternative for the community.

## 4 DETAILED CHEMICAL ABUNDANCES

In this work, to determine the chemical abundances of our sample, we implemented the Brussels automated code for characterizing high-accuracy spectra (BACCHUS; Masseron et al. 2016), on the reduced spectra of APOGEE DR17 (Abdurro'uf et al. 2022), using the `tonalli` atmospheric parameters determined in the previous section.

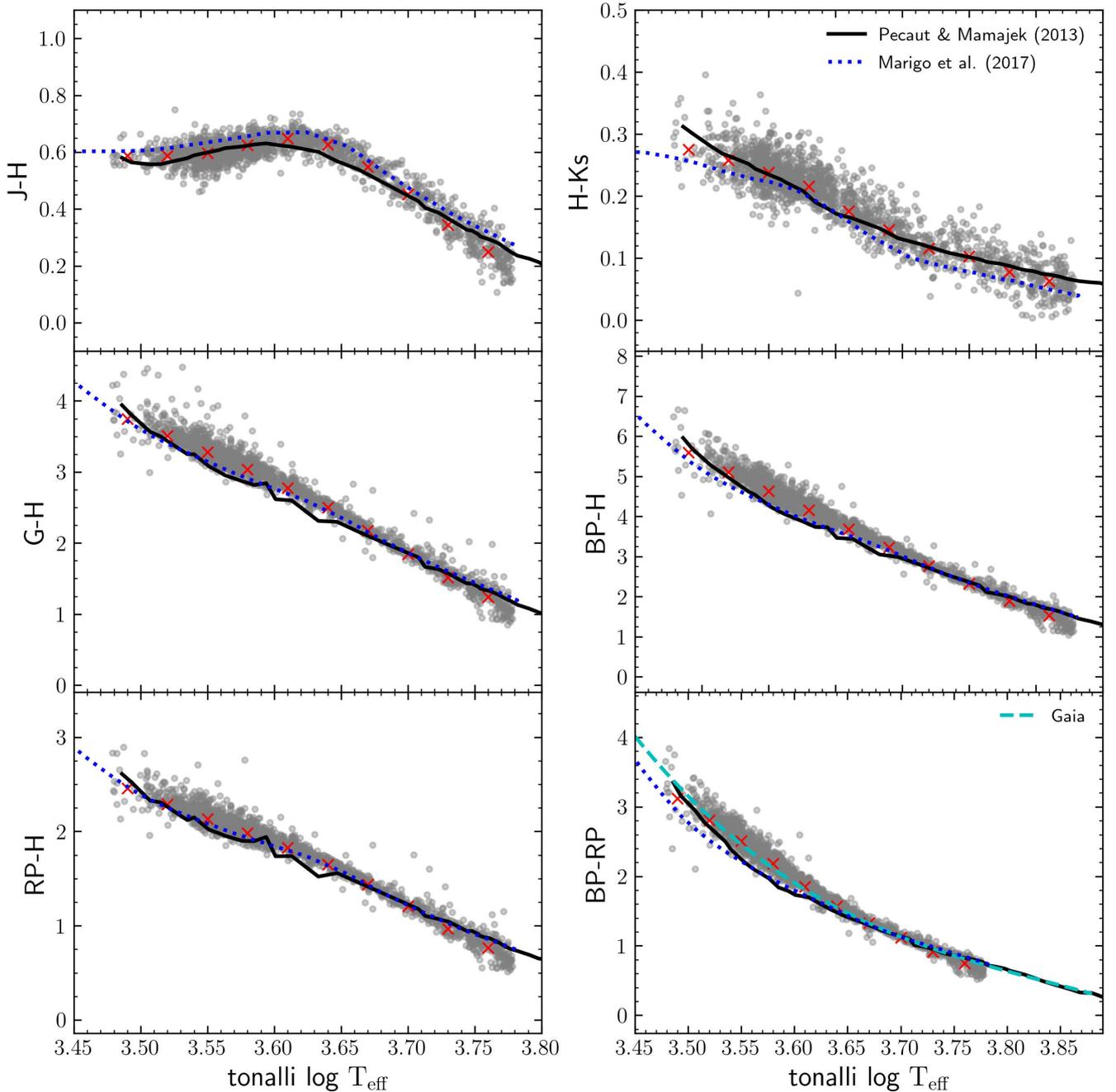

**Figure 6.** Gaia and 2MASS photometric colors as function of logarithmic tonalli $T_{eff}$. The black solid lines are the intrinsic colors by Pecaut & Mamajek (2013), the cyan dashed line is a polynomial fit to Gaia DR3 temperatures and colors, and the blue dotted lines are the theoretical colors of PARSEC (Marigo et al. 2017) for a $\log(Age) = 9.8$. The red crosses are the median color within the ten $T_{eff}$ binnes we split our sample and correspond to those reported in Table 2.

### 4.1 BACCHUS

BACCHUS computes on-the-fly synthetic spectra for a range of abundances and compares them to the observations on a line-by-line basis. The synthetic spectra in BACCHUS are calculated using the v15.1 of the 1D LTE Turbospectrum radiative transfer code (Plez 2012), a set of MARCS model atmospheres, specially built for APOGEE (Jönsson et al. 2020), the line list of Smith et al. (2021), and the molecular line lists for CO, CN, and OH of Li et al. (2015), Sneden et al. (2014), and Brooke et al. (2016) respectively. In addition to the line list and atmosphere models, Turbospectrum needs to be provided with (or compute) the $T_{eff}$, $\log g$, [M/H], [$\alpha$/Fe], microturbulence velocity, and a convolution parameter that includes the instrumental and rotation broadening to compute a synthetic spectrum. To derive chemical abundances, BACCHUS first normalizes the observed spectra, finding continuum points around the line of interest in a 30 Å window. After a sigma-clipping procedure is applied on those continuum points, a linear fit is performed to define a pseudo-continuum. Then, the normalized observed and synthetic spectra are compared within the spectral window through four different methods, namely, *chi2*, *syn*, *eqw*, and *int*.

Every method of BACCHUS has its advantages and disadvantages, and we suggest reading the respective documentation. Among all methods, the more balanced is *chi2*. The *chi2* method, that we used in this work, looks for the abundance that minimizes the differences between observed and synthetic spectra. BACCHUS flags the quality of the abundance estimations and produces plots for each analyzed line where the fitted models are compared to the observed spectrum, which can also be used to visually verify the goodness of the resulted abundances.

### 4.1.1 line list

We started our atomic line selection from a set of lines based on the linelist of Holtzman et al. (2018). This subset of lines has been used in different studies where the chemical abundances were derived with BACCHUS and APOGEE data (e.g., Masseron et al. 2019; Fernández-Trincado et al. 2021, 2022) and they were selected avoiding blended lines, hot pixels, and lines with poor telluric correction. We used this subset and obtained their atomic data information from the newer linelist of Smith et al. (2021). The line list of Smith et al. (2021) and its previous versions (e.g., Shetrone et al. 2015; Holtzman et al. 2018), was developed specially to be adopted by ASPCAP. The line list comprises literature values of theoretical and laboratory atomic information. The authors also derived and incorporated astrophysical oscillator strengths and damping parameters into the line list derived through Arcturus and solar APOGEE spectra.

We visually inspected the sub-set of lines to choose suitable lines for our study. We used the solar spectrum, reflected in the asteroid Vesta and 2 - 3 random spectra of stars with different ASPCAP $T_{eff}$ and $v \sin i$ to cover our $T_{eff}$ and $v \sin i$ ranges.

We first selected those lines deep enough to be measured in the solar spectra. We rejected lines with depths lower than ~1-2%. After that, and when possible, we selected isolated lines, rejecting those lines with evident blends or close neighboring lines. Once we have selected the best lines using the solar spectrum, we picked random stars with different $T_{eff}$ and $v \sin i$ to identify lines not suitable for all (or most) of the spectral types included in our sample. After our selection process, we kept 35 lines in total: 3 of Mg I, Al I, and Ca I, 10 of Si I, and 16 of Fe I. It is important to mention that in this study we are preferably selecting strong atomic lines. The basic information of the selected lines is reported in Table 3.

### 4.1.2 Chemical abundance computation

For each star in our sample, we obtained the epoch-combined 1D APOGEE spectrum[7] that we converted into air wavelengths using the Morton (1991) formulation.

We used this spectrum, the tonalli atmospheric parameters ($T_{eff}$, log $g$, [M/H], and [$\alpha$/Fe]), and the line list of Table 3 in BACCHUS to obtain the chemical abundances of Mg, Al, Si, Ca, and Fe. We left the microturbulence velocity and the convolution parameter of BACCHUS to vary freely. However, we initialized the convolution parameter to be equal to the tonalli $v \sin i$ value. The microturbulence is determined in BACCHUS by obtaining a null trend between the abundance

---

**Table 3.** Basic information of the 35 atomic lines used to determine the abundances with BACCHUS.

| air $\lambda$ (Å) | element | E (eV) | log gf | Van der Wals |
|---|---|---|---|---|
| 15740.716 | Mg I | 5.931 | -0.323 | -6.84 |
| 15748.988 | Mg I | 5.932 | 0.049 | -6.89 |
| 15765.842 | Mg I | 5.933 | 0.320 | -6.98 |
| 16718.990 | Al I | 4.085 | -0.0186 | -7.15 |
| 16750.539 | Al I | 4.087 | -0.233 | -7.22 |
| 16763.358 | Al I | 4.087 | -1.640 | -7.16 |
| 15376.831 | Si I | 6.223 | -0.701 | -7.00 |
| 15557.779 | Si I | 5.964 | -.0820 | -6.97 |
| 15884.454 | Si I | 5.954 | -0.945 | -7.11 |
| 15960.063 | Si I | 5.984 | 0.130 | -7.02 |
| 16060.009 | Si I | 5.954 | -0.452 | -7.07 |
| 16094.787 | Si I | 5.964 | -0.088 | -7.10 |
| 16215.670 | Si I | 5.954 | -0.565 | -7.21 |
| 16241.833 | Si I | 5.964 | -0.762 | -7.13 |
| 16680.770 | Si I | 5.984 | -0.138 | -7.24 |
| 16828.159 | Si I | 5.984 | -1.058 | -7.10 |
| 16150.763 | Ca I | 4.532 | -0.237 | -6.71 |
| 16157.364 | Ca I | 4.554 | -0.219 | -6.83 |
| 16197.075 | Ca I | 4.535 | 0.089 | -6.59 |
| 15207.526 | Fe I | 5.385 | 0.067 | -7.21 |
| 15244.974 | Fe I | 5.587 | -0.134 | -7.01 |
| 15621.654 | Fe I | 5.539 | 0.280 | -7.04 |
| 15631.950 | Fe I | 5.351 | -0.032 | -7.12 |
| 15662.013 | Fe I | 5.828 | 0.101 | -7.08 |
| 15723.586 | Fe I | 5.621 | -0.011 | -7.05 |
| 15904.324 | Fe I | 6.365 | -0.154 | -7.19 |
| 15920.642 | Fe I | 6.258 | -0.011 | -7.01 |
| 15980.726 | Fe I | 6.264 | 0.736 | -6.91 |
| 16125.899 | Fe I | 6.351 | 0.618 | -6.91 |
| 16284.769 | Fe I | 6.398 | 0.099 | -6.88 |
| 16316.320 | Fe I | 6.280 | 0.857 | -6.90 |
| 16517.223 | Fe I | 6.286 | 0.516 | -6.91 |
| 16561.765 | Fe I | 5.979 | 0.028 | -6.80 |
| 16645.874 | Fe I | 5.956 | -0.226 | -6.90 |
| 16665.482 | Fe I | 6.017 | -0.235 | -6.83 |

of Fe and equivalent widths. On the other hand, the convolution parameter is derived by matching the Fe abundance from the *int* and *eqw* methods.

BACCHUS produces an abundance file for each element, where it lists the computed absolute abundance, A(X)[8], for each method and each atomic line, accompanied by a quality flag.

Once the five element abundances are computed, we checked the abundance files generated by BACCHUS and we directly rejected those lines with flags different from 1, which is the flag that BACCHUS gives for a good fit, as well as those with SNR lower than 150. We set this high arbitrary value of SNR to ensure that we are taking into account only the best atomic lines for each observed spectrum. Finally, we did an iterative sigma clipping process on those abundances flagged as good fits until all individual abundances were within $2\sigma$ the mean value.

As a sanity check, we visually inspect the Fe abundance plots that BACCHUS produced for a sub-sample of 80 random stars. We visually chose to keep or discard a specific atomic line based on how well the models reproduce the observed spectra. Then, we computed the mean abundance value of these visually selected lines and compared them to those determined by the sigma-clipping process. We found a good agreement between both quantities, ensuring that both methods are consistent. In the following, we used the sigma-clipping procedure to compute the mean abundances.

---

[7] We used the apStar spectra downloaded from the SDSS-IV science archive web app (SAW)

[8] A(X) = log ($n_X/n_H$) + 12

**Table 4.** Solar mean abundances and their respective standard deviation obtained from the 500 Monte Carlo realizations. As a comparison, we also included the solar values obtained in two different studies.

| Element | this work | Asplund et al. (2005) | Lodders (2019) |
|---------|-----------|----------------------|----------------|
| Mg | 7.58±0.14 | 7.53±0.09 | 7.50±0.05 |
| Al | 6.43±0.10 | 6.37±0.06 | 6.41±0.03 |
| Si | 7.55±0.08 | 7.51±0.04 | 7.52±0.06 |
| Ca | 6.40±0.06 | 6.31±0.04 | 6.34±0.06 |
| Fe | 7.50±0.08 | 7.45±0.05 | 7.52±0.05 |

*4.1.3 Solar abundances*

To increase the consistency of our analysis, we derived, from the APOGEE-2 DR17 spectrum of Vesta, the stellar parameters as well as the solar abundances under the same circumstances and followed the same methodology as in our stellar sample. The solar abundances are important to characterize our abundance scale as they are used to obtain the [X/H] abundance ratios.

To get the solar abundances, we ran a Monte Carlo simulation taking into account the errors on the stellar parameters derived with `tonalli`. We determined 500 times the solar chemical abundances with `BACCHUS`, assigning each time a random set of parameters to the Vesta APOGEE DR17 spectrum. Such parameters are selected from randomly generated Gaussian distributions where the means are the `tonalli` parameters for Vesta: 5898 K, 4.7 dex, 0.03 dex, and −0.04 dex for $T_{\rm eff}$, $\log g$, [M/H], and [$\alpha$/Fe] respectively. The $\sigma$ of those Gaussians corresponds to the errors on such stellar parameters (123 K for $T_{\rm eff}$, 0.2 dex for $\log g$, 0.07 dex for [M/H], and 0.04 dex for [$\alpha$/Fe]).

Once `BACCHUS` computed the abundances, we applied our sigma-clipping procedure to find the mean abundance and its error (standard deviation) of the solar Mg, Al, Si, Ca, and Fe. At the end of the process, we had a posterior distribution of 500 mean abundances for each element. We used the central values of those posterior distributions as our solar reference abundance for each element. The standard deviation, on the other hand, gives us a sense of the error in our solar abundances.

The mean solar abundance and its standard deviation are reported in Table 4. The solar abundances we found here are in good agreement with different solar abundance scales (e.g., Grevesse & Sauval 1998; Lodders 2003; Asplund et al. 2005, 2009; Lodders 2019).

**4.2 Chemical abundance ratios**

We subtracted the corresponding solar value from our derived absolute abundances (see Table 4) to compute [X/H] abundance ratios[9].

The total error ($\sigma_{[X/H]}$) that we assigned to our [X/H] values is the quadrature sum of the uncertainties on the star and solar abundances. As mentioned before, we considered the standard deviation of the abundances of good-fitted atomic lines as the error in the abundance determination, while the error in our solar abundance is reported in Table 4.

[9] The [X/H] = A(X)$_{\rm star}$ − A(X)$_\odot$, where the A(X) are the `BACCHUS` abundances found

**Table 5.** Descriptive content of the Table of the stellar parameters and chemical abundances determined in this work. The full table is available in the electronic version of the paper and the VizieR database.

| Column name | Description |
|-------------|-------------|
| OBJ | 2MASS identification number |
| RA | Right Ascencion (J2000) |
| Dec | Declination (J2000) |
| t25 | Effective temperature at percentile 25 |
| t50 | Effective temperature at percentile 50 |
| t75 | Effective temperature at percentile 75 |
| g25 | Surface gravity at percentile 25 |
| g50 | Surface gravity at percentile 50 |
| g75 | Surface gravity at percentile 75 |
| m25 | Overall metallicity at percentile 25 |
| m50 | Overall metallicity at percentile 50 |
| m75 | Overall metallicity at percentile 75 |
| a25 | alpha elements abundance at percentile 25 |
| a50 | alpha elements abundance at percentile 50 |
| a75 | alpha elements abundance at percentile 75 |
| v25 | Projected rotational velocity at percentile 25 |
| v50 | Projected rotational velocity at percentile 50 |
| v75 | Projected rotational velocity at percentile 75 |
| Mg | [Mg/H] abundance |
| eMg | error on [Mg/H] abundance |
| lMg | number of lines used in abundance determination |
| Al | [Al/H] abundance |
| eAl | error on [Al/H] abundance |
| lAl | number of lines used in abundance determination |
| Si | [Si/H] abundance |
| eSi | error on [Si/H] abundance |
| lSi | number of lines used in abundance determination |
| Ca | [Ca/H] abundance |
| eCa | error on [Ca/H] abundance |
| lCa | number of lines used in abundance determination |
| Fe | [Fe/H] abundance |
| eFe | error on [Fe/H] abundance |
| lFe | number of lines used in abundance determination |

Our final [X/H] abundances and their errors are listed in Table 5. We did not measure the abundances for all the stars of our sample for two main reasons: i) the temperature of the star (too hot or cold) interferes in some cases to reliably measure some of the selected atomic lines. There might also be contamination by neighboring lines or by molecular features (especially at lower $T_{\rm eff}$), however, there is not a specific $T_{\rm eff}$ where the abundances were not measurable; and ii) there are cases where some spectral regions present data reduction issues, like the presence of cosmic rays, insufficient background extraction, among others. Those problems result in a lower SNR of one or more spectral lines, impeding us from meeting the strict SNR ≳ 150 cut we implemented. For the mentioned reasons, the number of abundance determinations varies for each element and star. We determined the abundance of the five elements simultaneously for 892 sources.

In Figure 7, we show our [X/H] abundance ratio as a function of our $T_{\rm eff}$ and $\log g$. We have binned our data into bins of 500 K and 0.30 dex in $T_{\rm eff}$ and $\log g$, respectively, to better understand our results. In the next subsections, we discuss each element in more detail.

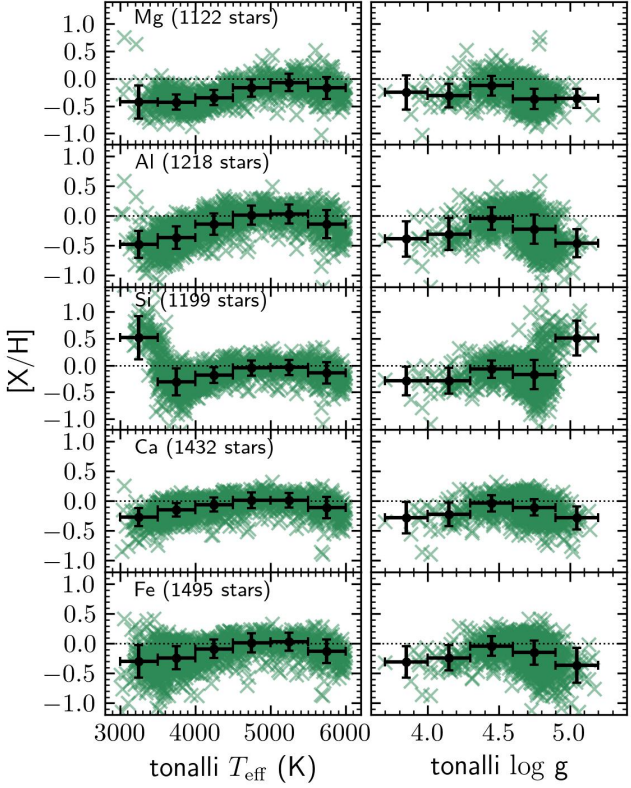

**Figure 7.** [X/H] abundance ratios as a function of tonalli's $T_{eff}$ (left panel) and $\log g$ (right panel). The black crosses are the same results but in bins of width of 500 K and 0.3 dex, respectively. The center of the crosses is the median [X/H], while the size of the y-axis cap represents the standard deviation of [X/H] within the bin and the x-axis cap is the size of the bin. The median errors on [X/H] are ∼0.10, while for $T_{eff}$ is ∼60 K.

*Iron*

The Fe is the element with the highest number of determinations in our analysis with 1495. In Figure 7, the median of the bins shows no trend of our [Fe/H] with $\log g$. Concerning $T_{eff}$, there is a slight trend being [Fe/H] higher for higher $T_{eff}$, however, the difference in the medians of the coolest and hottest bin is ∼0.17 dex. which is of the order of the dispersion in the data. Our sample has a mean [Fe/H] of -0.17±0.22 dex. If we consider just the G and K stars ($T_{eff} \geq 4000$ K), the mean increases up to -0.08±0.18 dex, which is in agreement with the value of Sousa et al. (2008) of -0.08±0.22. Sousa et al. (2008) determined the basic stellar parameters and [Fe/H] for 451 FGK stars observed in the HARPS GTO planet search programs.

In Figure 8 we show the difference between the ASPCAP DR16 calibrated abundances and our [X/H] values as a function of our $T_{eff}$ and [Fe/H] estimations. In appendix A, we included the same comparisons but for the DR17 values. The calibration performed by ASPCAP DR16 to their abundances is a zero-point shift to force stars with solar [M/H] in the solar neighborhood to have [X/M] = 0. The resulting shifts for most of the elements studied here are of the order of ±0.04 dex (see Table 4 of Jönsson et al. 2020), therefore, we compare our chemical abundances with the calibrated ASPCAP DR16 values.

The median difference (ASPCAP - tonalli) between abundances is about 0.03 dex. The abundance differences, in the left panel of Figure 8, do not show a trend with $T_{eff}$, and they are close to zero along the complete temperature range. Strictly speaking, the residuals have a lower dispersion (between 0.02 and 0.06 dex) for stars with a $T_{eff}$ between 4000 and 5500 K, meaning that our abundances agree better with ASPCAP DR16 in this temperature range. However, even for hotter or cooler temperatures the agreement is good enough as the median is around zero and the dispersion, in the worst case, is 0.2 dex.

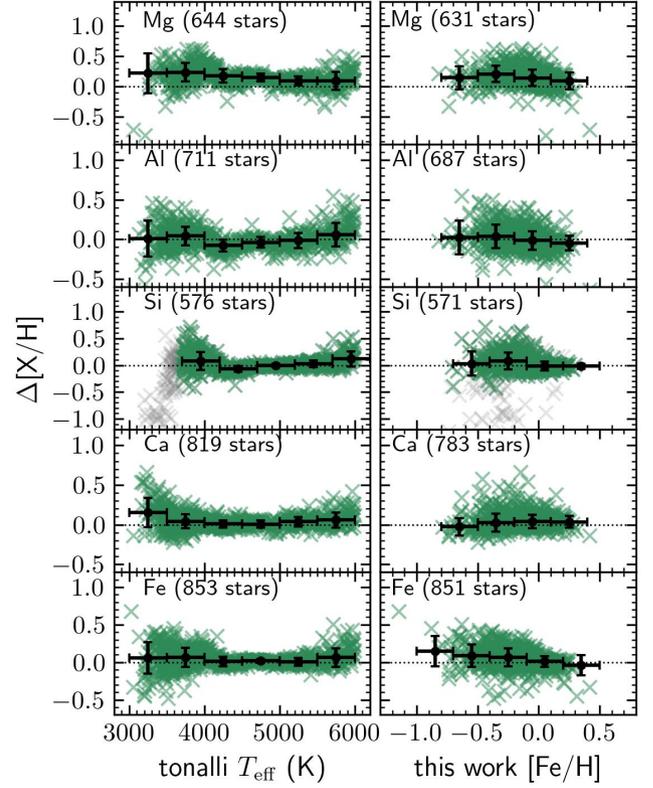

**Figure 8.** Difference between ASPCAP DR16 (Jönsson et al. 2020) calibrated chemical abundances and those determined in this work, as a function of our $T_{eff}$ (left panel) and [Fe/H] (right panel) estimations. The grey data in the Si panel are the results for stars with $T_{eff}$ lower than 3700 K. The black crosses are the same results but in bins of width of 500 K and 0.3 dex, respectively. The center of the crosses is the median Δ[X/H], while the size of the y-axis cap represents the standard deviation of Δ[X/H] within the bin and the x-axis cap represents the size of the bin.

On the other hand, the [Fe/H] residuals show a trend (see bottom right panel of Figure 8), where our [Fe/H] values are lower than those reported by ASPCAP DR16 for sub-solar iron abundances. In comparison, our [Fe/H] values are slightly higher than ASPCAP DR16 for solar or super-solar abundances. We found that for 489 stars our iron abundance is solar, i.e., [Fe/H] = 0.0±0.1 dex. Our [Fe/H] values have a dispersion of 0.05 dex, which is the same found in DR16 estimations for the same metallicity threshold.

*Silicon*

We determined the Si abundance for 1199 stars of our sample. We found a [Si/H] values distribution centered at -0.14±0.28 dex. However, as we can see in the third panel (top to bottom) of Figure 7, our Si abundances for the cooler stars ($T_{eff} \lesssim 3700$ K) are anomalously higher (up to 1.0 dex). A visual inspection of some of these stars shows that many of the transitions of Si become very complicated to measure at lower temperatures, for that reason, the Si abundance for the cool stars comes from just one line in about ~70% of the cases. Additionally, the median SNR of the cooler stars is lower than in hotter counterparts, due to the way APOGEE spectra are taken. A very detailed manual abundance determination should improve the Si values for these stars, but this is out of the scope of this paper.

Although some cool stars present [Si/H] consistent with the rest of our sample, we excluded from our analyses and conclusions the [Si/H] abundances for stars cooler than 3700 K, reducing to 1005 our Si determinations. We advise the reader to exercise caution using our Si abundances in this temperature range, which even for the cool stars are presented in Table 5.

Besides the mentioned behavior, our [Si/H] shows no obvious trend with $T_{eff}$, as a linear regression to the binned data has a slope close to zero. Regarding gravity, we see that those cool stars have also higher log $g$ values. Taking out from our analysis the cool stars, we did not find any trend between [Si/H] and log $g$.

Our [Si/H] are in excellent agreement with the DR16 calibrated values for G and K stars as shown in the third panel (top to bottom) of Figure 8. Similarly to [Fe/H] the highest dispersion in the differences of Figure 8 are for stars with temperatures between 3700 and 4000 K. We also did not find any considerable trend in the residuals of Si with [Fe/H] (slope of -0.08 dex). For the solar metallicity objects, we found a mean [Si/H] of -0.08 dex with a dispersion of 0.09 dex. The dispersion in our Si estimations is slightly higher than that found in DR16, which is 0.06 dex.

Neves et al. (2009) determined the abundance of 12 elements (including Mg, Al, Si, and Ca) in the sample of 451 FGK stars of Sousa et al. (2008). Of these 451 stars, 360 are classified as part of the thin disk and have a $T_{eff}$ lower than 6000 K. Since our sample comprises stars highly likely from the thin disk (see Section 2), we compare, in this and following sections, our results just with the 360 thin disk stars of Neves et al. (2009). If we considered only G and K stars (≥ 4000 K), the mean [Si/H] is -0.12±0.17 dex, which is lower but still in agreement within the uncertainties with the reported value of -0.04±0.22 dex by Neves et al. (2009).

In the left panel of Figure 9 we show our abundances relative to iron as a function of `tonalli` temperature. We also include in the same Figure the respective DR16 abundance ratio as a comparison. As we did in Figures 7 and 8, we binned our data in steps of 500 K to better understand the behavior of the abundance ratio. We found for our [Si/Fe] median values close to zero in stars with $T_{eff} \gtrsim 4200$ K, comparable to DR16 results in the same temperature range. DR16 estimations have a slight trend, where their [Si/Fe] begins in zero and then starts to be negative up to $T_{eff}$ of about 4250±250 K where the value of [Si/Fe] peaks at −0.12 dex. Then, their [Si/Fe] constantly increases up to values of −0.02 dex, for the hotter stars. Our [Si/Fe] values for the coolest stars (3700 ≤ $T_{eff}$ ≤ 4200 K) have a large dispersion of 0.14 dex, and the median is -0.12 dex, similar to the lowest median [Si/Fe] value of DR16.

*Calcium*

We determined the Ca abundance for 1432 stars. The [Ca/H] distribution of our complete sample has a mean value of -0.12±0.15 dex. If we take into consideration only stars with $T_{eff}$ ≥ 4000 K (G and K stars), our mean value for [Ca/H] is -0.06±0.15 dex, which is in excellent agreement with the value determined by Neves et al. (2009) of -0.04±0.18 dex. The fourth panels (top to bottom) of Figure 7, did not show any trend with our $T_{eff}$ nor log $g$ values. When we compare our Ca abundances with ASPCAP DR16 (Figure 8), we found that for stars with temperatures lower than 3500 K our estimations are lower, but the residuals did not show any considerable trend with $T_{eff}$ or [Fe/H]. It is clear, that our [Ca/H] are in good agreement with the DR16 calibrated values. For solar metallicities, we found a mean [Ca/H] of -0.03±0.07 dex in agreement, within the uncertainties, with the 0.01±0.06 dex for the stars of our sample with DR16 solar metallicity values.

Our [Ca/Fe] in Figure 9 shows a small dispersion for stars hotter than 4000 K. The median [Ca/Fe] for the hotter stars is between -0.01 and 0.03 dex with dispersions between 0.04 and 0.06 dex. These results are very similar to DR16, where their median and dispersions span comparable ranges to ours. For stars with $T_{eff}$ lower than 4000 K, our [Ca/Fe] values are more dispersed and with median values slightly higher than zero. However, this behavior is also present in the DR16 catalog, pointing out the difficulty of obtaining precise abundances in cool stars.

*Aluminum*

The Al were determined in 1218 stars of our sample. The [Al/H] values have a mean value of -0.22±0.25 dex. If we only consider the G and K ($T_{eff}$ ≥ 4000 K) stars the [Al/H] mean is -0.08±0.20 dex in agreement with the 0.01±0.20 dex found by Neves et al. (2009). From the second (top to bottom) panels of Figure 7 we see a hint that there is a trend for stars with a temperature lower than 5000 K, being the median [Al/H] lower for lower $T_{eff}$. Interestingly the dispersion within the temperature bins is about ~0.19 dex in all of them. When we compare our [Al/H] with our log $g$ the result is as in all the previously mentioned elements, i.e. there is no trend between both quantities. The comparison between our Al estimations and those of ASPCAP DR16 (see Figure 8) shows a slope close to zero when the residuals are a function of our $T_{eff}$ value. When we compare our residuals with our [Fe/H] values we found that the median of the bins is close to zero, meaning that we have a good agreement with the Al estimations of DR16. A linear regression to the binned residuals shows a slope close to -0.09dex, which is similar to the values found for Ca and Si, however, the dispersion within the bins also increases as [Fe/H] decreases. We found a median difference between DR16 and our Al abundances of -0.01±0.14 dex. Regarding the solar metallicity stars, we found a mean [Al/H] of -0.06±0.13, while for DR16 estimations the mean is -0.12±0.12 dex.

Regarding our [Al/Fe] abundances (Figure 9), we found median values close to zero for stars with temperatures between 4000 and 6000 K. Compared to DR16, we see that our [Al/Fe] behaves very similarly, where the median abundance ratio is constant and close to zero for hot stars while for cooler objects ($T_{eff} < 4000$ K) the value start to become more and more negative. The fact that we share the same [Al/Fe] pattern with DR16 might be related to the use of similar atomic data and model atmosphere in their computation.

*Magnesium*

We determined the Mg abundance in 1122 sources of our sample. The [Mg/H] distribution is centered at -0.30±0.20 dex. If we take into consideration only stars with $T_{eff}$ ≥ 4000 K (G and K stars), our mean value for [Mg/H] is -0.22±0.20 dex, which is lower than the value of -0.03±0.21 dex found by Neves et al. (2009), but it still agrees within the uncertainties. The dispersion we found in the [Mg/H] is comparable to that found by Neves et al. (2009). Our [Mg/H] as a function of $T_{eff}$ present an interesting behavior, it seems that our [Mg/H] have a decreasing trend for stars with temperatures between 5500 and 3500 K, then the abundances flatten for the coolest stars in our sample, this might be because the number of stars in the coolest bin is significantly lower than in the others (42 against hundreds of stars). The medians of the bins are close to each other when comparing our [Mg/H] with log $g$. We found a median difference of 0.16 dex when comparing our Mg abundances with those of ASPCAP DR16. There is a slight trend in the residuals of the abundances with tonalli $T_{eff}$, but the slope is too low and close to zero. A similar behavior of the residuals is shown with our [Fe/H] estimations (see Figure 8).

For the solar metallicity stars, we found a mean [Mg/H] of -0.20±0.14, while for the DR16 abundances is -0.05±0.05 dex. Although it is clear that our mean [Mg/H] for the solar metallicity stars is offset concerning the DR16 values, the means still agree within the uncertainties.

For our [Mg/Fe] (right panel of Figure 9) we found a trend with $T_{eff}$, where the abundance ratio increases as the $T_{eff}$ does. The median [Mg/Fe] of the temperature bins span a range from -0.26 to -0.05 dex and the dispersion is between 0.06 and 0.34 dex. On the contrary DR16 presents (right panel of Figure 9) a flat [Mg/Fe] - $T_{eff}$ plot, with median values close to zero.

*The α elements distribution*

The Mg, Si, and Ca are $\alpha$-elements. The alpha elements are mostly produced in core-collapse supernovas in a short time (3-20 Myr), while the Fe-peak elements are produced in type Ia supernovas in a longer time (0.1-20 Gyr). Due to the differences in the time scale of production of the alpha and Fe-peak elements, the [$\alpha$/Fe] abundance ratio is indicative of the chemical evolution of the Galaxy.

We compute the [$\alpha$/Fe] ratio of our sample by adding our [Mg/Fe], [Si/Fe], and [Ca/Fe] abundances. We found a median of -0.22±0.18 dex for the complete sample. If we focus just on stars with solar metallicities ([Fe/H] = 0.0±0.1 dex) we found a median [$\alpha$/Fe] of -0.25±0.16 dex. We notice that our [$\alpha$/Fe] values display a trend with $T_{eff}$ as median values for G, K, and M stars are -0.13±0.10 dex, -0.30±0.09 dex, and -0.45±0.17 dex, respectively, at solar metallicities.

However, This trend is driven by [Mg/Fe], which as discussed before, has lower abundances for cooler stars (see Figure 9). If we discard our [Mg/Fe] abundance ratios from the computation of [$\alpha$/Fe] and limit us to stars hotter than 4000 K the trend between [$\alpha$/Fe] and $T_{eff}$ is removed. Now the median values are -0.04±0.4 dex and -0.07±0.5 dex, for G and K stars respectively. These results are comparable with the expected value of [$\alpha$/Fe]=0.0 for solar metallicities.

As just described and as it is summarized in Table 6, the five [X/H] abundance distributions of our sample have sub-solar mean values consistent with previous determinations for the solar neighborhood where studies found [X/H] distributions centered at solar or slightly sub-solar values (e.g., Sousa et al. 2008; Neves et al. 2009; Sousa et al. 2011). The differences between our results and those of Neves et al. (2009) and Sousa et al. (2008) might be explained by using a different solar normalization, methodology, wavelength interval, and/or spectral resolution. On the other hand, the difference found for Mg is more significant to be fully explained by the previously mentioned causes as they could only account for about ~0.06 dex (Hinkel et al. 2014). In the case of Mg, we should not discard non-LTE effects or inaccurate atomic information of the line list.

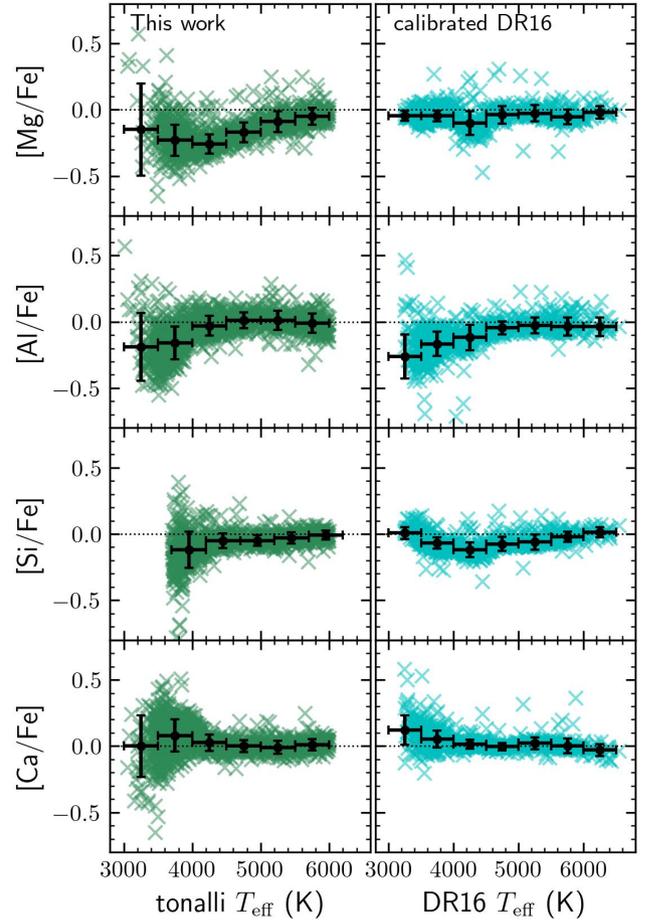

**Figure 9.** [X/Fe] abundance ratios as a function of temperature. In the left panel, we show the results obtained here, while in the right panel are the calibrated DR16 values. The black crosses are the same results but in bins with a width of 500 K. The center of the crosses is the median [X/Fe], while the size of the y-axis cap represents the standard deviation of [X/Fe] within the bin and the x-axis cap represents the size of the bin.

The comparison of chemical abundances between different studies is a complicated process, as there is no such thing as absolute abundances. Their determination depends on various aspects, such as line lists, model atmosphere, standard solar composition, and techniques. The fact that we found good agreement with the DR16 calibrated abundances implies that our methodology produces chemical abundances consistent with previous estimations, with the advantage that we are not applying any posterior calibration to obtain the atmospheric parameters nor a zero-point shift to our abundances. In the appendix A3, we present the analogous Figures 8 and 9 for DR17 abundances. In general, DR17 abundances have slightly improved with respect to DR16, and thus the agreement with our abundance estimations also has improved.

We found that our abundance values for the M stars ($T_{eff}$ < 4000 K) are, on average, lower than the hotter stars (see Table 6). It is important to consider that stars of different spectral types (G, K, and M) do not necessarily share the same chemical abundances and stellar ages, although they are close main-sequence stars. In the particular case of the M stars, the difference in abundance with the G and K stars, and the fact that they have lower abundance ratios, might be related to issues in the model atmospheres, as to date models are still far from reproduced the cool star spectra at the same level as for earlier spectral types.

If we consider that DR17 supersedes the DR16 catalog and we consider it as a good atmospheric parameters and chemical abundances collection, our techniques are producing better atmospheric parameters and chemical abundances consistent with such catalogs. This is very important to our study as we are validating that our methodology is working properly and we can apply it in other stellar populations, such as pre-main sequence stars, where there might not be well-characterized samples to compare with. This study also helps to increase the number of stars with atmospheric parameters and chemical abundance values determined homogeneously, which might be useful as a training set or reference sample in machine learning techniques/studies of our Galaxy or even in other astronomical fields.

## 5 SUMMARY

We present a methodology to determine atmospheric parameters and elemental chemical abundances from APOGEE-2 high-resolution infrared spectra. We determined $T_{\rm eff}$, $\log g$, [M/H], [$\alpha$/Fe], $v \sin i$ with tonalli (Adame et al. in prep) in a sample of over 1600 main-sequence nearby stars (d $\leq$ 100 pc). We compute the Galactic space velocities to ensure that our stars belong to the thin disk. Our sample comprises 1653 stars, that according to our $T_{\rm eff}$ values, they might have spectral types between G and M (6000 K $\leq T_{\rm eff} \leq$ 3000 K).

Our atmospheric parameters are as good as previous estimations as comparison with ASPCAP DR16 (Jönsson et al. 2020), DR17 (Abdurro'uf et al. 2022), ANet III (Sizemore et al. 2024), and GAIA DR3 (Gaia Collaboration et al. 2023a) show. We established an optical-infrared color–temperature relation, which is an excellent alternative to other existing relations (e.g., Pecaut & Mamajek 2013), and it can be used for any G, K, or M main-sequence star.

We use the atmospheric parameters determined with tonalli in BACCHUS (Masseron et al. 2016) to compute the chemical abundance of Mg, Al, Si, Ca, and Fe. The comparison between our values and those of ASPCAP DR16 shows an acceptable agreement between studies; however, our estimations seem to be lower. We found that our sample has a mean abundance centered around a sub-solar value for the five elements, which is in agreement with previous studies of the solar neighborhood (Sousa et al. 2008; Neves et al. 2009; Sousa et al. 2011). With the present study, we are increasing the number of stars with computed chemical abundances. For example, we increase up to 50% the number of stars with Al abundance of ASPCAP DR17 for sources with $T_{\rm eff} >$ 4000 K, while we were able to determine the Al abundance in more than 500 cooler stars ($T_{\rm eff} \leq$ 4000 K), something that ASPCAP DR17 lacks.

The atmospheric parameters and the chemical abundances determined in this work are an excellent collection that can be used in a diversity of studies involving the solar neighborhood. Moreover, our sample is also of great importance as a well-characterized training set for data-driven methods (e.g., neuronal networks), something that has increased in popularity in recent years and will undoubtedly be the future of the characterization of large stellar samples. Additionally, the methodology presented here might be easily implemented in pre-main sequence objects.

## ACKNOWLEDGEMENTS

R. L-V and L. A. acknowledge support from CONAHCyT through a postdoctoral fellowship within the program "Estancias Posdoctorales por México". J.G.F-T gratefully acknowledges the grant support provided by Proyecto Fondecyt Iniciación No. 11220340, and from the Joint Committee ESO-Government of Chile 2021 (ORP 023/2021) and 2023 (ORP 062/2023). SV gratefully acknowledges the support provided by Fondecyt regular n. 1220264 and by the ANID BASAL projects FB210003. LC acknowledges support from the grant IG-100622. (DGAPA-PAPIIT, UNAM). The authors acknowledge support from projects CONAHCYT CB2018-A1-S-9754, CONAHCYT CF86372 and UNAM PAPIIT IG101723. Funding for the Sloan Digital Sky Survey IV has been provided by the Alfred P. Sloan Foundation, the U.S. Department of Energy Office of Science, and the Participating Institutions. SDSS-IV acknowledges support and resources from the Center for High-Performance Computing at the University of Utah. The SDSS website is www.sdss.org. SDSS-IV is managed by the Astrophysical Research Consortium for the Participating Institutions of the SDSS Collaboration including the Brazilian Participation Group, the Carnegie Institution for Science, Carnegie Mellon University, the Chilean Participation Group, the French Participation Group, Harvard-Smithsonian Center for Astrophysics, Instituto de Astrofísica de Canarias, The Johns Hopkins University, Kavli Institute for the Physics and Mathematics of the Universe (IPMU) / University of Tokyo, Lawrence Berkeley National Laboratory, Leibniz Institut für Astrophysik Potsdam (AIP), Max-Planck-Institut für Astronomie (MPIA Heidelberg), Max- Planck-Institut für Astrophysik (MPA Garching), Max-Planck-Institut für Extraterrestrische Physik (MPE), National Astronomical Observatories of China, New Mexico State University, New York University, University of Notre Dame, Observatário Nacional / MCTI, The Ohio State University, Pennsylvania State University, Shanghai Astronomical Observatory, United Kingdom Participation Group, Universidad Nacional Autónoma de México, University of Arizona, University of Colorado Boulder, University of Oxford, University of Portsmouth, University of Utah, University of Virginia, University of Washington, University of Wisconsin, Vanderbilt University, and Yale University. This research has made use of the SIMBAD database, operated at CDS, Strasbourg, France.

## DATA AVAILABILITY

Table 5 is available in the online supplementary material of the paper and the VizieR database.

**Table 6.** Mean abundances for our sample in three $T_{eff}$ ranges, corresponding roughly to spectral types G, K, and M, respectively. In columns 2 to 6, we show in parenthesis the number of stars considered in the computation of the mean and standard deviation. While the number in the parenthesis in column 1, is the total number of stars of our sample within the $T_{eff}$ range.

| $T_{eff}$ range | [Mg/H] | [Al/H] | [Si/H] | [Ca/H] | [Fe/H] |
| --- | --- | --- | --- | --- | --- |
| 5000 – 6000 K (309) | -0.14±0.19 (305) | -0.08±0.22 (306) | -0.11±0.19 (309) | -0.07±0.17 (309) | -0.08±0.19 (309) |
| 4000 – 5000 K (422) | -0.28±0.17 (399) | -0.08±0.18 (393) | -0.13±0.16 (417) | -0.05±0.13 (410) | -0.07±0.16 (416) |
| < 4000 K (922) | -0.42±0.16 (418) | -0.40±0.20 (519) | -0.34±0.19 (279) | -0.18±0.13 (713) | -0.26±0.2 (770) |

## APPENDIX A: ASPCAP DR16 AND DR17 COMPARATIVE PLOTS

The ASPCAP pipeline in its DR16 and DR17 versions implemented a posterior calibration to its $T_{eff}$ and log $g$ values and to their chemical abundances. In the main manuscript, we just considered the calibrated DR16 estimations to avoid confusion with the different sets of parameters and versions. In this appendix, we present the related comparisons between our parameters and those determinations of DR17. We also compare our values with the DR16 *spec* determinations.

### A1 DR17 Kiel Diagrams

In Figure A1 we present the Kiel diagram of our sample for both *spec* and *calibrated* ASPCAP DR17 $T_{eff}$ and log $g$ values. Similarly to *spec* DR16 Kiel diagram (left panel in Fig. 2) many stars fall outside of the expected region of the diagram for $T_{eff} \lesssim 4800$ K. The posterior calibration performed in DR17 (right panel of Figure A1) is much better than that of DR16, and it produces log $g$ values with less dispersion.

### A2  Comparisons of DR16 and DR17 *spec* $T_{eff}$ and log $g$

Figure A2 compares our $T_{eff}$ values to the *spec* estimations of DR16 and DR17. The median residuals (literature - `tonalli`) for DR16 is −85 K and −157 K for DR17. In general, the agreement between temperatures is good for both data releases, being slightly better for DR16. We also found that the shape of the residuals is similar: between ∼3500 and ∼4800 K the ASPCAP temperatures are cooler than those of `tonalli`, then for hotter stars ($T_{eff}$> 5000 K) the agreement improves up to ∼5800-5900 K, which relate with the applicability limits of `tonalli`.

Similarly to $T_{eff}$, we compare in Figure A3 our log $g$ values with those *spec* estimations within ASPCAP DR16 and DR17. We found that in DR16, DR17, and `tonalli` the log $g$ values fell between 4 and 5 dex. Although this, the one-to-one comparison shows a trend in the residuals (see lower panels of Figure A3) of both data releases. For `tonalli` low log $g$ values the *spec* DR16 estimations are higher (up to ∼0.5 dex), while the DR16 values are lower, on average more than 0.5 dex, for the higher `tonalli` log $g$. This behavior is similar for *spec* DR17 estimations. It is clear from the left panel of Figure 4 and Figure A4 that the posterior calibration performed on the *spec* log $g$ values of DR16 and DR17 improves the agreement with `tonalli` determinations. Still, the agreement is much better with the calibrated DR16 values than with the DR17, as there is a persistent trend in the residuals of DR17 (see lower panel of Figure A4). This means that tour log $g$ values are at least as good as the calibrated DR16 and DR17 results with the advantage of using an informed prior rather than a posterior calibration.

### A3  ASPCAP DR17 chemical abundances comparison

This section compares our chemical abundances with those reported in the ASPCAP DR17 catalog (Abdurro'uf et al. 2022). The left panels of Figure A5 show the differences (DR17 - this work) of the chemical abundances as a function of our $T_{eff}$ and [Fe/H]. In general, we see that our abundance estimations agree with DR17 for Mg, Si, Ca, and Fe, as the median differences are close to zero. The residual of the Al abundances for stars with $T_{eff} \gtrsim$ 5500 K exhibit a trend where the higher the temperature the higher the differences. Additionally, DR17 does not report Al abundances for stars cooler than ∼ 4500 K, while our results included Al abundance for more than 700 stars in this temperature range.

In the right panels of Figure A5, we show the same differences but now as a function of our [Fe/H] value. We see the strongest trend with [Fe/H] in the residuals of Al. We found a slope of -0.32 dex in the residuals of Al. For Si and Fe the slopes are similar and close to ∼ −0.13 dex. Finally, the residuals of Mg and Ca depict a slope of -0.06 and 0.04 dex, respectively.

In Figure A6, we show our [X/Fe] as a function of our $T_{eff}$ in the left panel, while in the right panel, we present the [X/Fe] of DR17. The DR17 [X/Fe] present less dispersion than their DR16 counterpart values and the median values are close to zero. In general, the agreement between DR16 and DR17 is good, even the DR17 [Si/Fe] improves with respect to DR16. However, the DR17 [Mg/Fe] ratios are slightly worse than their counterpart of DR16. The main difference between catalogs is that in DR17 the number of objects with determinations increased, but a down is that DR17 removed determinations for lower temperatures stars ($T_{eff} \lesssim$ 3500 K).

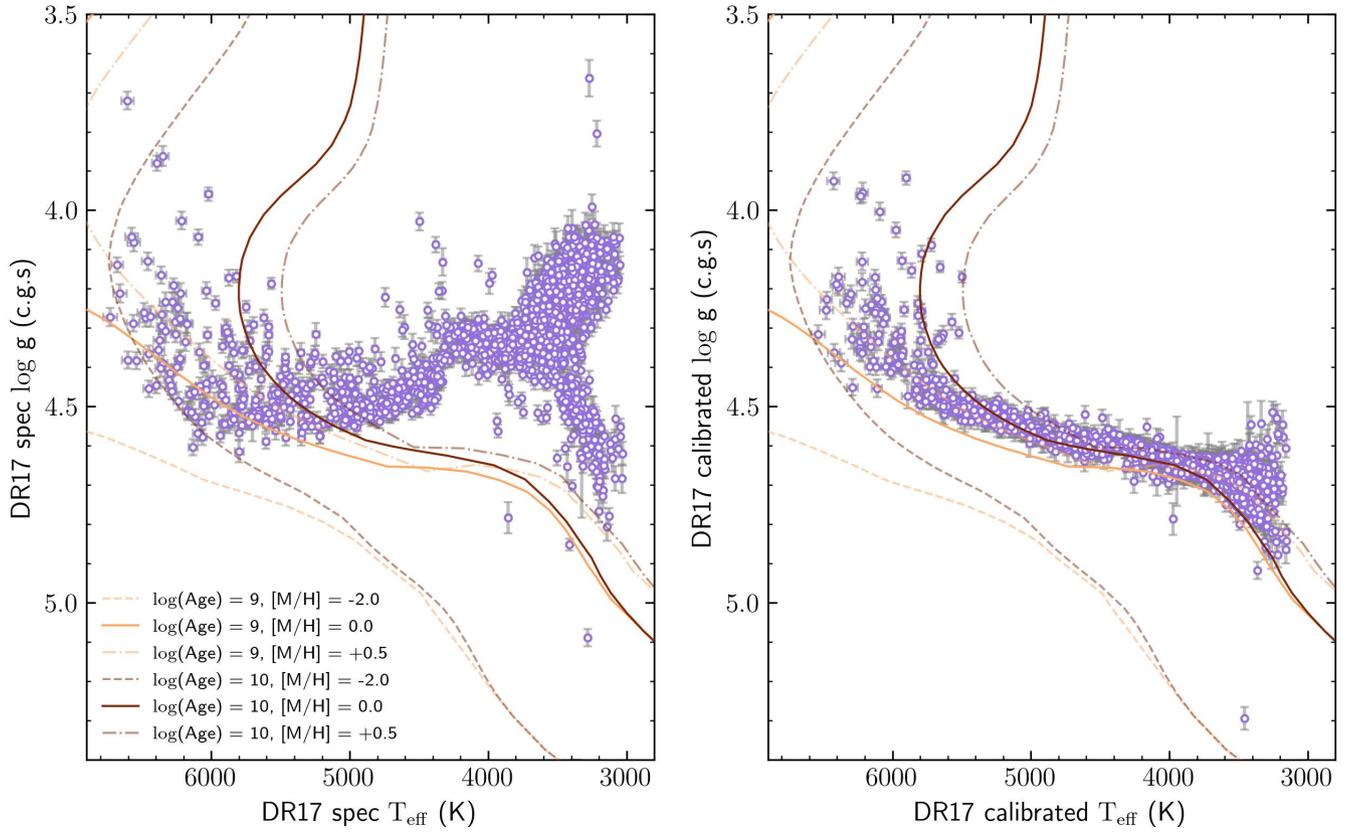

**Figure A1.** Kiel diagram for the 1653 stars of our sample and the atmospheric parameters determined by ASPCAP DR17. On the left panel, we show the spec values, which are the raw parameters determined by the pipeline, while on the right panel, there are the calibrated values. We also added some PARSEC isochrones for reference.

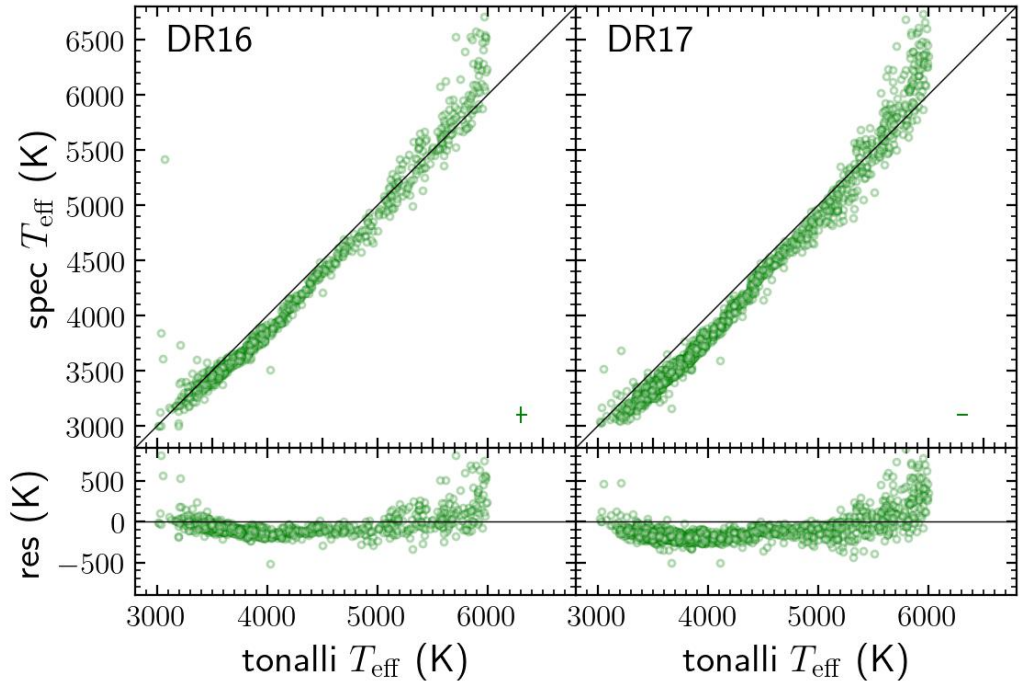

**Figure A2.** Comparison of the `tonalli` T$_{\rm eff}$ with those *spec* estimations of ASPCAP DR16 (upper left panel) and DR17 (upper right panel). The lower panels are the residuals between the literature and our values. The solid line represents the one-to-one relation. In the bottom right corner of each upper panel are the median errors.

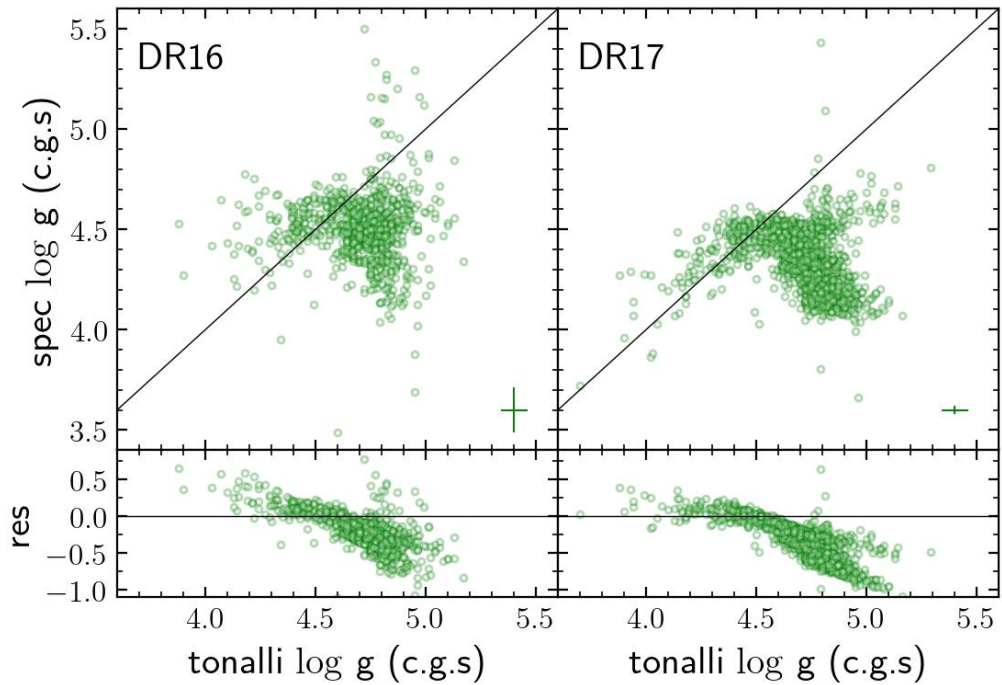

**Figure A3.** Comparison of the `tonalli` log *g* with those *spec* estimations of ASPCAP DR16 (upper left panel) and DR17 (upper right panel). The lower panels are the residuals between the literature and our values. The solid line represents the one-to-one relation. In the bottom right corner of each upper panel are the median errors.

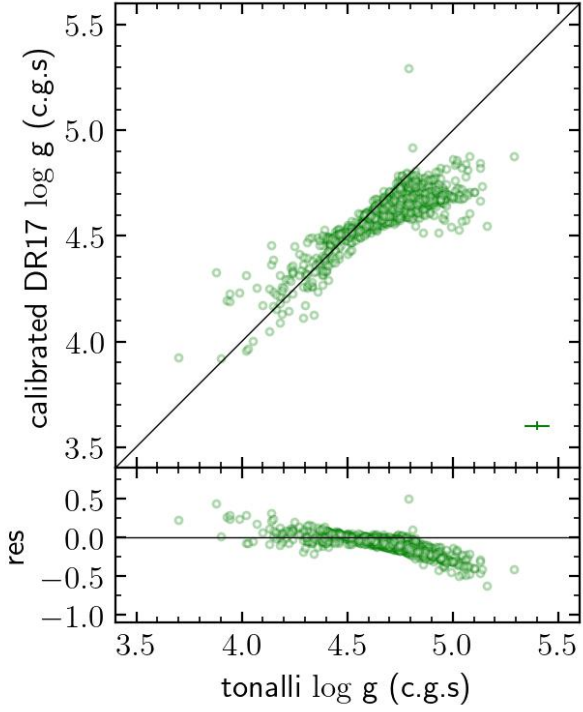

**Figure A4.** Comparison between the log *g* values determined in this study and the calibrated DR17 values. The lower panel shows the residuals (DR17 - `tonalli`) of both gravities. The solid line represents the one-to-one relations while the median errors are in the bottom right corner.

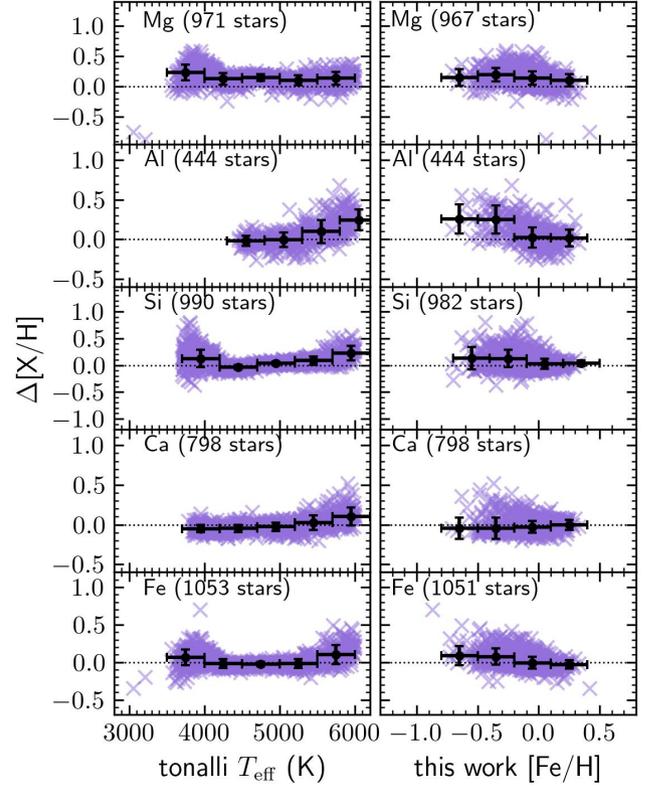

**Figure A5.** Difference between ASPCAP DR17 (Abdurro'uf et al. 2022) calibrated chemical abundances and those determined in this work, as a function of our $T_{\mathrm{eff}}$ (left panel) and [Fe/H] (right panel) estimations. The grey data in the Si panel are the results for stars with $T_{\mathrm{eff}}$ lower than 3700 K. The black crosses are the same results but in bins of width of 500 K and 0.3 dex, respectively. The center of the crosses is the median $\Delta$[X/H], while the size of the y-axis cap represents the standard deviation of $\Delta$[X/H] within the bin and the x-axis cap represents the size of the bin.

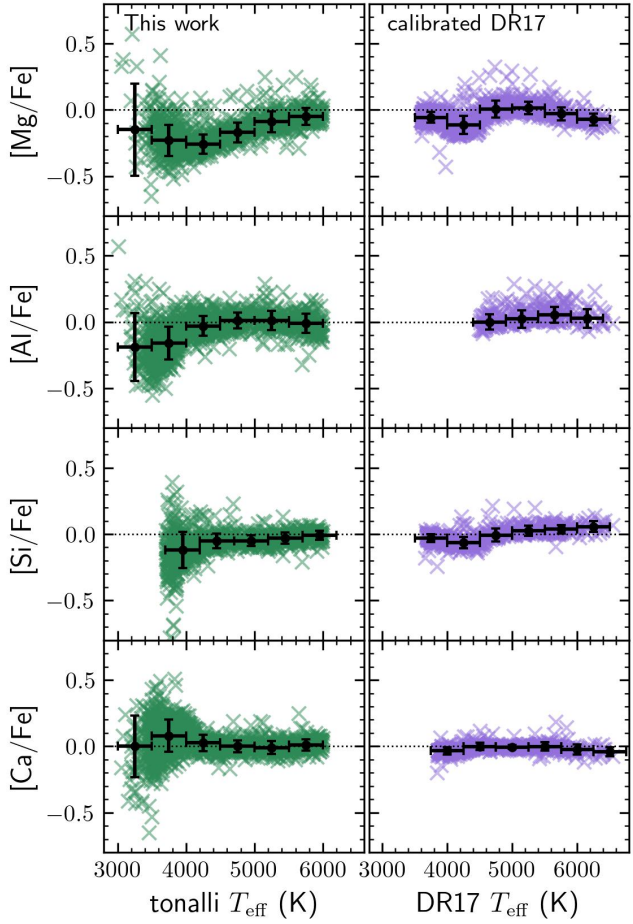

**Figure A6.** [X/Fe] abundance ratios as a function of temperature. In the left panel, we show the results obtained here, while in the right panel are the calibrated DR17 values (Abdurro'uf et al. 2022). The black crosses are the same results but in bins of width of 500 K. The center of the crosses is the median [X/Fe], while the size of the y-axis cap represents the standard deviation of [X/Fe] within the bin and the x-axis cap represents the size of the bin